\documentclass[12pt]{article}

\usepackage{amsmath}
\usepackage{amssymb}
\usepackage{epsfig}
\usepackage{graphicx}
\makeatletter
\renewcommand{\fnum@table}{\textbf{\tablename~\thetable}}
\renewcommand{\fnum@figure}{\textbf{\figurename~\thefigure}}
\makeatother

\newcounter{myenumi}

\renewcommand{\themyenumi}{\roman{myenumi}}
{\end{list}}

\newlength{\myem}
\newcounter{mysubequation}[equation]

\makeatletter
\renewcommand{\section}{\@startsection{section}{1}{0em}{-\baselineskip}%
{\baselineskip}{\normalfont\large\bfseries}}
\renewcommand{\subsection}%
{\@startsection{subsection}{2}{0em}{-0.7\baselineskip}%
{0.7\baselineskip}{\normalfont\bfseries}}
\makeatother

\newcommand{\bea}{\begin{eqnarray*}}
\newcommand{\eea}{\end{eqnarray*}}

\newcommand{\deltacp}{\delta_\mathrm{CP}}

\newcommand{\GeV}{\,\mathrm{GeV}}

\newcommand{\eV}{\,\mathrm{eV}}

\newcommand{\reu}{{\nu_e\rightarrow\nu_\mu}}
\newcommand{\reub}{{\bar{\nu}_e\rightarrow\bar{\nu}_\mu}}
\newcommand{\ruu}{{\nu_\mu\rightarrow\nu_\mu}}
\newcommand{\ruub}{{\bar{\nu}_\mu\rightarrow\bar{\nu}_\mu}}

\newcommand{\dm}[1]{{\Delta m^2_{#1}}}

\newcommand{\szt}[1]{\sin^22\theta_{#1}}

\newcommand{\Psl}{sub-leading}
\newcommand{\Pssl}{sub-sub-leading}
\newcommand{\PSl}{Sub-leading}
\newcommand{\PSsl}{Sub-sub-leading}

\DeclareMathOperator{\im}{Im}
\DeclareMathOperator{\re}{Re}

\begin{document}

%%%%%%%%%%%%%%%%%%%%%%%%%%%%%%%%%%%%%%%%%%%%%%%%%%%%%%%%%%%%%%%%%%%%%
%%%%                     Title-page                              %%%%
%%%%%%%%%%%%%%%%%%%%%%%%%%%%%%%%%%%%%%%%%%%%%%%%%%%%%%%%%%%%%%%%%%%%%

\begin{titlepage}

% the footnote symbols are only redefined for the title page !
\renewcommand{\thefootnote}{\alph{footnote}}

\ \vspace*{-3.cm}
\begin{flushright}
TUM-HEP-414/01\\
MPI-PhT/01-13\\
hep-ph/0105071
\end{flushright}

\vspace*{0.5cm}

\renewcommand{\thefootnote}{\fnsymbol{footnote}}
\setcounter{footnote}{-1}

{\begin{center} 
{\large\bf Systematic Exploration of the Neutrino Factory Parameter Space}\\[2mm]
{\large\bf including Errors and 
        Correlations$^*$\footnote{\hspace*{-1.6mm}$^*$Work supported by 
        "Sonderforschungsbereich 375 f\"ur Astro-Teilchenphysik" der 
        Deutschen Forschungsgemeinschaft.}}
\end{center}}
\renewcommand{\thefootnote}{\alph{footnote}}

\vspace*{.8cm}
\vspace*{.3cm}
{\begin{center} {\large{\sc
                M.~Freund\footnote[1]{\makebox[1.cm]{Email:}
                Martin.Freund@ph.tum.de},~  
                P.~Huber\footnote[2]{\makebox[1.cm]{Email:}
                Patrick.Huber@ph.tum.de}~and~
                M.~Lindner\footnote[3]{\makebox[1.cm]{Email:}
                lindner@ph.tum.de}~
                }}
\end{center}}
\vspace*{0cm}
{\it 
\begin{center}  
        
\footnotemark[1]${}^,$\footnotemark[2]${}^,$\footnotemark[3]%  
       Theoretische Physik, Physik Department, 
       Technische Universit\"at M\"unchen,\\
       James--Franck--Strasse, D--85748 Garching, Germany

\footnotemark[2]% 
       Max-Planck-Institut f\"ur Physik, Postfach 401212, 
       D--80805 M\"unchen, Germany 

\end{center}}

\vspace*{1.5cm}

{\Large \bf 
\begin{center} Abstract \end{center}  }
We discuss in a systematic way the extraction of neutrino masses, 
mixing angles and leptonic CP violation at neutrino factories.
Compared to previous studies we put a special emphasis on 
improved statistical methods and on the multidimensional nature 
of the combined fits of the $\reu$, $\reub$ appearance and $\ruu$, 
$\ruub$ disappearance channels. Uncertainties of all involved 
parameters and statistical errors are included. We find previously 
ignored correlations in the multidimensional parameter space,
leading to modifications in the physics reach, which amount 
in some cases to one order of magnitude. Including proper
statistical errors we determine for all parameters the improved 
sensitivity limits for various baselines, beam energies, neutrino 
fluxes and detector masses. Our results allow a comparison of the 
physics potential for different choices of baseline and beam energy 
with regard to all involved parameters. In addition we discuss in 
more detail the problem of parameter degeneracies in measurements of 
$\deltacp$.

\vspace*{.5cm}

\end{titlepage}

\newpage

\renewcommand{\thefootnote}{\arabic{footnote}}
\setcounter{footnote}{0}

%%%%%%%%%%%%%%%%%%%%%%%%%%%%%%%%%%%%%%%%%%%%%%%%%%%%%%%%%%%%%%%%%%%%%
%                     Introduction                                  %
%%%%%%%%%%%%%%%%%%%%%%%%%%%%%%%%%%%%%%%%%%%%%%%%%%%%%%%%%%%%%%%%%%%%%

\section{Introduction \label{sec:SEC-intro}}

The potential to measure neutrino masses, mixings, matter 
effects and leptonic CP violation at neutrino factories has 
already been studied for different setups 
(see e.g. \cite{Albright:2000xi,FLPR, FHL}). 
Among the main points discussed are the achievable precision for
the oscillation parameters, the optimal baseline and the best muon 
energy, the search for matter effects, the search for CP violation
and specifically, whether measurements of the CP phase are possible. 
The results of such studies depend, however, strongly on the chosen
experimental setup and the assumed physics parameters, and the 
combined total parameter space is not easily overlooked. 

A common strategy to deal with this situation is to discuss only 
one or two parameters simultaneously, while the remaining 
parameters are set to ``standard values'' (like the muon
flux or $\sin^2 2\theta_{23}=1$). This is a good method 
for first estimates of the physics capabilities of such experiments.
Statistical errors computed in this way are, however, underestimated
since they miss possible correlations with the parameters which are
hold fixed. This can lead to quite severe errors, especially close 
to the sensitivity limits, where a strong interplay between several 
parameters exists. Similar problems arise in a two step analysis, 
where the disappearance channel is analyzed with \Psl~corrections
ignored, and where the extracted leading parameters are used 
for the extraction of \Psl~parameters.

We present in this work an event rate analysis based on a statistical 
method, which treats all parameters on equal footing instead of arbitrarily 
selecting specific parameter subsets. The best method would be to 
perform full six parameter fits to the simulated event rates, but 
this is not feasible, since the required computing time for the 
exploration of the parameter space and the parameter dependencies 
is enormous. We adopt therefore a somewhat simplified, but still 
general method which is based on the calculation of all two dimensional 
slices through the fit manifold. This method is related to the covariance 
matrix method and it automatically includes all two parameter 
correlations. We apply this method to the analysis of all involved 
physics parameters as a function of the machine parameters. 
In particular, we give improved sensitivity limits for measurements 
of the mixing angle $\theta_{13}$, matter effects and CP violation. 
There, the impact of the improved statistical treatment is most evident: 
Sensitivity limits can be deteriorated by up to one order of magnitude 
by correlations with \Psl~parameters. We demonstrate furthermore how 
input from other experiments (like KamLAND \cite{KamLAND}) helps 
to improve the results obtained from neutrino factories. Finally, 
we study the possibility of combining two distinct baselines and we 
present results, which help to determine the optimal baseline and 
the best muon energy. 

The paper is organized as follows: First we present in 
section~\ref{sec:formalism} analytic formulae for the relevant 
oscillation probabilities in vacuum and in matter, which allows 
to understand many numerical results qualitatively. In  
section~\ref{sec:classification} we provide a classification 
scheme of the parameter space, which is important, since it 
provides an overview of how we analyze and discuss the complex 
parameter space in detail in section~\ref{sec:results}. The 
framework which we use for our numerical simulation of event
rates has been discussed in detail in earlier works. We explain 
this framework therefore only briefly in section~\ref{sec:numerics} 
and we give references to earlier studies. The statistical 
methods which we used in our study are described in detail in 
section~\ref{sec:statistics}. The results are given in 
section~\ref{sec:results}, where we follow the classification 
scheme introduced in sec.~\ref{sec:classification}. 
Finally, in section~\ref{sec:conclusion} we conclude.

%%%%%%%%%%%%%%%%%%%%%%%%%%%%%%%%%%%%%%%%%%%%%%%%%%%%%%%%%%%%%
\section{Framework}
%%%%%%%%%%%%%%%%%%%%%%%%%%%%%%%%%%%%%%%%%%%%%%%%%%%%%%%%%%%%%

%%%%%%%%%%%%%%%%%%%%%%%%%%%%%%%%%%%%%%%%%%%%%%%%%%%%%%%%%%%%%
\subsection{Oscillation probabilities in vacuum and in matter}
%%%%%%%%%%%%%%%%%%%%%%%%%%%%%%%%%%%%%%%%%%%%%%%%%%%%%%%%%%%%%
\label{sec:formalism}

We assume standard three neutrino mixing with the leptonic 
mixing matrix $U$, which coincides with the standard 
parameterization of the quark mixing matrix \cite{PDG}.
The probabilities which describe in vacuum flavour oscillations 
between an arbitrary number of neutrinos are
\begin{equation}
P_{lm}= \delta_{lm} 
- 4\sum_{k>h}\re(J^{e_m e_l}_{kh})\sin^2\Delta_{kh} 
- 2\sum_{k>h}\im(J^{e_m e_l}_{kh})\sin(2\Delta_{kh})~,
\label{Plm}
\end{equation}
where $J^{e_i e_j}_{kh} := U_{e_i\nu_k} 
U_{\nu_k e_j}^\dagger U_{e_j\nu_h} U_{\nu_h e_i}^\dagger$, 
$\Delta_{kh} := \dm{kh} L/(4 E_\nu)$ and $\Delta:=\Delta_{31}$.
Matter effects lead to sizable corrections of these probabilities.
The numerical results shown in section~\ref{sec:results} are based 
on event rates, calculated from the full oscillation formulae, with 
the matter potential included. As usual, it is assumed that only 
muons can be detected and a number of the results can be understood 
analytically from the corresponding probabilities $P(\ruu)$ and $P(\reu)$.
The full expressions are, however, quite lengthy and do not allow 
much insight. We provide therefore simplified expressions by expanding 
these probabilities in the small mass hierarchy parameter 
$\alpha := \dm{12}/\dm{31}$. Interesting CP- and $\dm{12}$-effects 
will only occur for bi-maximal mixing and if $\alpha$ is not too 
small, i.e. for the LMA-MSW solution of the solar neutrino deficit. 
We focus therefore on the currently favoured LMA-MSW 
solution of the solar neutrino deficit \cite{Bahcall:2001hv}.
The expansion in $\alpha$ is a good approximation as long as 
the oscillation governed by the solar $\dm{21}$ is small compared 
to the leading atmospheric oscillation, i.e.~ $\alpha{\Delta} \lesssim 1$. 
This translates into a upper bound for the baseline: 
\begin{equation}
\label{LLIMIT}
L \lesssim 8000\,\mathrm{km} \left(\frac{E_\nu}{\GeV}\right)  
\left(\frac{10^{-4}\eV^2}{\dm{21}}\right)  ~.
\end{equation}
Up to order $\alpha^2$ one obtains for the vacuum appearance and 
disappearance probabilities:
\begin{eqnarray}
\label{PROBVACUUM}
P(\nu_e \rightarrow \nu_\mu) &\approx& \sin^2 \theta_{23} 
\sin^2 2\theta_{13} \sin^2 {\Delta} \nonumber \\
&\pm&  \alpha\; \sin\deltacp \cos\theta_{13} \sin 2\theta_{12} 
\sin 2\theta_{13} \sin 2\theta_{23}
\sin^3{\Delta} \nonumber \\
&+&  \alpha\; \cos\deltacp \cos\theta_{13} \sin 2\theta_{12} 
\sin 2\theta_{13} \sin 2\theta_{23}
 \cos {\Delta} \sin^2 {\Delta} \nonumber  \\
&+& \alpha^2 \cos^2 \theta_{23} \sin^2 2\theta_{12} \sin^2 {\Delta} \\
\label{DISPROBVACUUM}
P(\nu_\mu \rightarrow \nu_\mu) &\approx& 1 - \cos^2 \theta_{13} 
\sin^2 2\theta_{23} \sin^2 {\Delta} \nonumber \\
&+&  2 \alpha  \cos^2 \theta_{13} \cos^2 \theta_{12} 
\sin^2 2\theta_{23} {\Delta} \cos{\Delta} 
\end{eqnarray}
The numerical magnitude of the different terms in the expansion is 
affected by different powers of the small mixing angle $\theta_{13}$. 
All terms of the expansion can be written in the form 
$\alpha^{n_\alpha} \theta_{13}^{n_\theta}$ and 
for $\alpha \simeq \theta_{13}$ the numerical size of each term is 
roughly controlled by $n=n_\alpha + n_\theta$. 
This amounts effectively to a reordering of the expansion
in eq.~(\ref{PROBVACUUM}), where all terms have $n=2$.
Note that all terms proportional to $\alpha$ have also a 
$\sin 2\theta_{13}$ factor and are thus typically of the same 
magnitude as the $\alpha^2$-term. Higher orders of the expansion 
in $\alpha$ are, however, always suppressed relative to these leading 
terms, since $n_\theta \geq 0$. The precise magnitude of the terms
which are in the sense of this reordering of the same magnitude
depends of course on the parameter values, especially on the 
size of $\alpha$ compared to $\theta_{13}$. For $\theta_{13}$ close 
to the present upper bound ($\sin^2 2\theta_{13} \approx 0.1$), 
the first term of eq.~(\ref{PROBVACUUM}) is, for example, dominating 
and the last term which is proportional to $\alpha^2$ is tiny and 
can be ignored. For smaller values of $\theta_{13}$, all four 
terms have approximately the same importance and all of them 
have to be considered for analytical explanations. The last 
$\alpha^2$-term in eq.~(\ref{PROBVACUUM}) can only become dominating
for extremely tiny values of $\theta_{13}$.

The CP phase $\deltacp$ produces only in the appearance channel significant 
effects. This can be seen from eq.~(\ref{DISPROBVACUUM}), where $\deltacp$
does not show up, while $\deltacp$ is contained in eq.~(\ref{PROBVACUUM}).
Both terms in eq.~(\ref{PROBVACUUM}), which contain information on the 
CP phase are, however, suppressed by the mass hierarchy. The precise 
value of $\dm{21}$ has thus considerable impact on the magnitude of 
CP-violating effects. There is a profound difference between seeing 
CP-violating effects directly in an experiment and optimally measuring 
the CP phase $\deltacp$ on the other hand. The point is that the term 
in (eq.~\ref{PROBVACUUM}) which is proportional to $\sin \deltacp$ 
changes sign when anti-neutrinos are considered and violates CP 
explicitely. The CP phase enters, however, also in the term 
which contains $\cos\deltacp$, which does not violate CP explicitely, 
but still can and should be used as an important lever arm to extract 
$\deltacp$ from measurements. We treat the CP phase $\deltacp$ in our 
analysis therefore exactly in the same way as all other parameters,
i.e. we fit rates to the full equations including the $\cos\deltacp$-term
and constrain the parameter space of $\deltacp$. Our numerical 
analysis also includes matter effects naturally, and the procedure 
has therefore the important advantage that it maximally exploits the 
information on $\deltacp$ contained in the appearance 
channels\footnote{Note that seeing explicit CP violation is not 
easy in the presence of matter. In vacuum an asymmetry between 
neutrino and anti-neutrino appearance rates would be a clear 
signal for CP violation, but such asymmetries arise in matter 
also from MSW effect, such that both effects must be carefully
separated.}.

For baselines above some hundred kilometers, one must include 
also the matter potential felt by neutrinos passing through 
the Earth. This opens the interesting possibility to observe MSW-effects
\cite{Wolfenstein:1978ue,Wolfenstein:1979ni,Mikheev:1985gs,
Mikheev:1986wj} in the appearance channel, while the disappearance 
probability $P(\nu_\mu \rightarrow \nu_\mu)$ is only marginally affected 
by matter effects. We include these effects in our numerical analysis and
for an analytic discussion we are interested in a reliable and traceable 
expression for $P(\nu_e \rightarrow \nu_\mu)$ in matter of constant average
density. A simple approximative result which is valid for small 
values of $\theta_{13}$ was derived in \cite{FREUND}, where also
expressions for larger values of $\theta_{13}$ are given.
Similar formulas were also derived in \cite{CERVERA,FLPR}.
\begin{eqnarray}
\label{eq:PROBMATTER}
P(\nu_e \rightarrow \nu_\mu) &\approx&  \sin^2 \theta_{23} 
\sin^2 2\theta_{13} \frac{\sin^2[(\hat{A}-1){\Delta}]}{(1-\hat{A})^2} 
\nonumber \\
&\pm&   \alpha\; \sin\deltacp \cos\theta_{13} \sin 2\theta_{12} 
\sin 2\theta_{13} \sin 2\theta_{23}
\sin({\Delta})~\frac{\sin(\hat{A}{\Delta})}{\hat{A}}
\frac{\sin[(1-\hat{A}){\Delta}]}{(1-\hat{A})}
\nonumber  \\
&+&   \alpha\; \cos\deltacp \cos\theta_{13} \sin 2\theta_{12} 
\sin 2\theta_{13} \sin 2\theta_{23}
\cos({\Delta})~\frac{\sin(\hat{A}{\Delta})}{\hat{A}}
\frac{\sin[(1-\hat{A}){\Delta}]} {(1-\hat{A})}
 \nonumber  \\
&+&  \alpha^2 \cos^2 \theta_{23} \sin^2 2\theta_{12}
\frac{\sin^2(\hat{A}{\Delta})}{\hat{A}^2} ~.
\end{eqnarray}
Here, $\hat{A} = A/\dm{31} = 2 V E_\nu/\dm{31}$ and 
$V = \sqrt{2} G_F n_e$ where $G_F$ is the Fermi coupling 
constant and $n_e$ the electron density of the involved 
matter profile. The expressions show at first sight that 
in the limit of small baselines, the vacuum result 
(eq.~\ref{PROBVACUUM}) is recovered.

%%%%%%%%%%%%%%%%%%%%%%%%%%%%%%%%%%%%%%%%%%%%%%%%%%%%%%%%%%%%%
\subsection{Classification of oscillation parameters}
%%%%%%%%%%%%%%%%%%%%%%%%%%%%%%%%%%%%%%%%%%%%%%%%%%%%%%%%%%%%%
\label{sec:classification}

The full three neutrino oscillation formulae eq.~(\ref{Plm})
have in general a rather complex parameter structure. All
mass splittings and mixing parameters, namely $\dm{31}$, 
$\dm{21}$, $\theta_{12}$, $\theta_{13}$, $\theta_{23}$ and 
$\deltacp$ appear in the relevant transition probabilities 
even after the expansion in $\alpha$ of section~\ref{sec:formalism}.
The expansion allows, however, to make use of two experimentally 
justified facts: First, the hierarchy of the neutrino mass 
splittings, i.e. $\alpha = \dm{12}/\dm{31} \ll 1$, which allows 
to identify hierarchy suppressed ``small $\dm{}$ effects''. 
Second, the mixing angle $\theta_{13}$ is small with 
$\sin^2 2\theta_{13} < 0.1$ \cite{Apollonio:1999ae}, leading
to further suppression factors as outlined in section~\ref{sec:formalism}.
The smallness of these two parameters allows to classify the 
parameter space with the help of the analytic formulae in the 
scheme below, which is useful for reasons of structure and 
clarity. Our results are, however, based on a full numerical
calculation, which essentially confirm this classification, 
but also fails in some places, as we will see later in this work.

{$\bullet$ \bf Leading parameters:}\\
For $\dm{21} = 0$ and $\theta_{13} =  0$, there are no 
transitions in the $\reu$ appearance channel. The disappearance 
probability reduces to the two neutrino case 
$P(\ruu) = 1 - \sin^2 2\theta_{23} \sin^2 {\Delta}$ being controlled 
by $\theta_{23}$, $\dm{31}$, which we call {\em leading parameters}. 
These parameters have already been 
measured by atmospheric neutrino experiments 
\cite{Scholberg:1999ar,Fukuda:2000np,Ambrosio:1998wu} and
will be determined better by conventional long 
baseline experiments \cite{Nakamura:2000uu,MINOS,CNGS1,CNGS2}. 
A neutrino factory will allow precision measurements of 
$\theta_{23}$ and $\dm{31}$ and the result will be limited mainly by 
systematical errors. The measurement of these parameters will
be dominated by the unsuppressed rates in the $\ruu$ and $\ruub$
disappearance channels, which does not rely on excellent charge 
identification capabilities for secondary muons \footnote{See 
e.g.~\cite{FHL} for a discussion of this problem.}. The differential 
event rate distribution allows precise fits of the energy
spectrum and the question is only how good $\theta_{23}$ and 
$\dm{31}$ can be measured for certain detector and neutrino 
factory parameters, i.e. what the optimal beam energy and 
baseline are in this context. It will also be interesting to see
how the accuracy for the leading parameters is modified for
baselines which optimize the sensitivity to $\sin^22\theta_{13}$ 
or CP violation. This will be addressed in section~\ref{sec:LEADING}.

{$\bullet$ \bf \PSl~parameters:}\\
For $\theta_{13} \neq 0$ and $\dm{21} = 0$, the first term in the 
appearance probabilities 
eq.~(\ref{PROBVACUUM}) or in matter eq.~(\ref{eq:PROBMATTER}) 
becomes non-zero: $P(\reu) = \sin^2 \theta_{23} \sin^2 2\theta_{13} 
\sin^2((\hat{A}-1){\Delta})/(\hat{A}-1)^2$. The appearance channel 
depends then via $\hat{A}$ on the sign of $\dm{31}$ 
\cite{FLPR,FHL,Barger:1999fs}. In addition to the leading 
parameters, the analysis depends at this level also on 
$\theta_{13}$ and $\mathrm{sgn}\dm{31}$, which we call 
{\em \Psl~parameters}. Whether it is possible 
to determine these \Psl~parameters depends crucially on the 
value of $\theta_{13}$, and the sensitivity limit below which 
no effects from $\theta_{13}$ can be measured will be studied 
in detail. Matter effects have so far not been measured 
and an experimental test of MSW-effects \cite{Wolfenstein:1978ue,
Wolfenstein:1979ni,Mikheev:1985gs,Mikheev:1986wj} will be 
possible if $\theta_{13}$ is large enough. The measurement 
of $\theta_{13}$ and the search for matter effects~\cite{FLPR,FHL,
Albright:2000xi,CERVERA} are thus important 
topics for the physics program of a neutrino factory, which do not 
depend on $\dm{21}$ and $\theta_{12}$ being in the LMA-MSW range.
Measurements of rates in the appearance channel depend, however,
crucially on the capability to identify the charges of the 
secondary muons very reliably \cite{FHL}, which requires 
improved detector technology. Sensitivity limits to $\theta_{13}$ 
and statistical errors are studied in section~\ref{sec:SUBLEADING},
where also baseline and beam energy optimization are discussed.

{$\bullet$ \bf \PSsl~parameters:}\\
Finally, for $\dm{21}\neq 0$ and $\theta_{13}\neq 0$, effects due 
to the small solar mass squared splitting are added and the 
remaining three parameters appear in the oscillation formulae: 
$\dm{21}$, $\theta_{12}$ and the CP phase $\deltacp$. 
Measuring leptonic CP violation is an exciting possibility for
neutrino factories. In order to obtain sufficient rates, this 
requires, however, that the LMA-MSW region is the correct 
solution to the solar neutrino problem. One can see immediately 
from eq.~(\ref{PROBVACUUM}) that $\dm{21}$ (i.e. $\alpha$) and 
$\theta_{13}$ are the crucial parameters with determine the 
absolute and relative strength of the CP-violating effects. 
The foreseeable experiments are often not too far from 
the sensitivity limit. The limits on those two parameters, 
which will be discussed in section~\ref{sec:SUBSUBLEADING},
are therefore among the most important points of
this study, providing a better understanding how to search 
for CP violation and how $\deltacp$ can be extracted.
We will also see in section~\ref{sec:SUBSUBLEADING} that 
$\theta_{12}$ and $\dm{21}$ can be measured only very 
poorly in neutrino factory experiments. To a good 
approximation only the product $\dm{21}\sin 2\theta_{21}$
can be determined, which can be expected already from 
the the appearance probability in eq.~(\ref{PROBVACUUM}).
Thus we expect and will see that $\dm{21}$ and $\theta_{12}$ 
are highly correlated. Results obtained from neutrino factories  
cannot compete with 
results expected from the long baseline reactor experiment 
KamLAND~\cite{BARGER}. It is thus very important to study 
how external input on $\dm{21}$ and $\theta_{12}$ 
(e.g. from  KamLAND) helps to improve the extraction of
CP-violating effects at a neutrino factory. The related 
questions about the optimal baseline and beam energy will 
be discussed in detail. As already mentioned above, the 
effects from the CP phase $\deltacp$ are, like the matter 
effects, best accessible in the appearance channels 
$\reu$ and $\reub$ which rely strongly on excellent charge 
identification capabilities of the detector.

Note that the extraction of both the \Psl~and \Pssl~parameters 
is essentially based on moderate event rates, which can be 
estimated from the oscillation probability at a mean energy.
Unlike the disappearance channels, there is thus essentially
no information in the energy spectrum. Note also that the above 
classification in leading, \Psl~and \Pssl~effects is also useful 
from a practical point of view. The experimental abilities of a 
neutrino factory will realistically develop from an initial, low 
flux setup, which is able to measure the leading parameters
only, to a high luminosity machine, capable of measuring 
\Psl~and probably \Pssl~effects. This will allow in a good 
approximation an extraction of parameters at the relevant 
level, where the remaining parameters of the level below 
act essentially as unknown variables.

%%%%%%%%%%%%%%%%%%%%%%%%%%%%%%%%%%%%%%%%%%%%%%%%%%%%%%%%%%%%%
\subsection{Event rate analysis of neutrino factory experiments}
\label{sec:numerics}
%%%%%%%%%%%%%%%%%%%%%%%%%%%%%%%%%%%%%%%%%%%%%%%%%%%%%%%%%%%%%

Our numerical analysis (see ref.~\cite{FHL} for details) is based 
on simulated event rates, which are calculated using neutrino 
flux profiles, full three neutrino oscillation probabilities 
in matter, a realistic Earth matter profile, neutrino cross 
sections, detector mass, threshold and energy resolution.  
Backgrounds and experimental uncertainties are not taken into 
account.  The $\nu_e-$, $\bar{\nu}_e-$, $\nu_\mu-$ and 
$\bar{\nu}_\mu-$fluxes produced by a symmetrically operated 
neutrino factory of given muon energy and polarization can 
easily be derived from the muon flux and the kinematics of muon 
decay. The transition probabilities are calculated numerically 
by propagating the neutrino states through a realistic Earth 
density profile. The resulting muon event rates are then computed 
by folding the neutrino flux, the transition probabilities and 
the cross sections with the energy resolution function of the 
detector. The assumed muon detection threshold is $4\GeV$.
The energy resolution of the detector is approximated by a 
Gaussian resolution function with $\sigma=0.1 E$, which 
gives an energy resolution $\Delta E/E=10\%$. This value 
resembles the energy resolution of MONOLITH~\cite{Agafonova:2000xm} 
for $\nu_{\mu}$ charged current events.

With this method, event rates are calculated for the 
$\reu$, $\reub$ appearance and $\ruu$, $\ruub$ disappearance
transitions with 20 energy bins in each channel. We assume 
perfect muon charge separation, i.e. that the different 
appearance and disappearance channels can be well separated. 
This requires for the appearance channels excellent charge
identification capabilities of the detector. Insufficient
charge identification would reduce the physics reach of such 
an experiment as pointed out in~\cite{FHL}. The type of detector 
assumed for this study is a magnetized iron detector with a
mass of 10~kt. With $2\cdot 10^{20}$ muon decays per year 
we obtain a standard ``luminosity'' of $N_\mu \, m_\mathrm{kt} 
= 2\cdot 10^{21} \,\mathrm{kt}\,\mathrm{year}$. We consider,
however, also variations of the parameter $N_\mu \, m_\mathrm{kt}$. 
Thus, different fluxes like an initial $10^{19}$ muon decays per 
year or higher fluxes (or detector masses) are considered.  
The details of the statistical analysis are explained in 
chapter \ref{sec:statistics}. The analysis uses simultaneously 
the neutrino and anti-neutrino channels of both the disappearance 
and appearance rate vectors, leading in all cases to optimal 
results. The method works of course also in cases where the 
results are dominated by one channel. Note, that $N_\mu$ is the 
sum of muons of both polarities, which means that the parameter 
$N_\mu\,m_\mathrm{kt}$ normalizes the sum of the neutrino and 
the anti-neutrino channels. Further details of the simulations
can be found in ref.~\cite{FHL}.

%%%%%%%%%%%%%%%%%%%%%%%%%%%%%%%%%%%%%%%%%%%%%%%%%%%%%%%%%%%%%
\section{Statistical methods}\label{sec:statistics}
%%%%%%%%%%%%%%%%%%%%%%%%%%%%%%%%%%%%%%%%%%%%%%%%%%%%%%%%%%%%%
 
There are a number of non-trivial aspects in neutrino factory
studies which require suitable statistical methods. 
We describe therefore in this section in more detail 
the problems and the methods used in this work. We also compare 
our method with other previously used methods.

The aim of the analysis is to obtain statistically reliable statements 
about possible measurements of $\dm{21},\,\dm{31},\,\theta_{12},\,
\theta_{13},\,\theta_{23}$ and $\deltacp$. A numerical scan of 
the full six dimensional input parameter with a subsequent extraction
of the input parameters is an enormous numerical task. Therefore,
up to now most studies performed only two parameter fits to evaluate
the physics reach of a neutrino factory~\cite{CERVERA,FHL,
DeRujula:1998hd,Barger:1999fs,Donini:1999jc,Dick:1999ed,Barger:2000ax,
Barger:2000cp,Albright:2000xi}. Often two parameters of the full 
parameter set, which were thought to be most relevant, were selected 
and a fit to this two parameters was performed using a $\chi^2$-method. 
All other parameters were at the same time fixed to their
best fit (i.e. input) values. Typically Gauss or Poisson 
$\chi^2$-functions have been used, i.e.
\begin{eqnarray}
\mathrm{Gauss}\quad\chi^2&=&
\sum_{i=1}^{b}\frac{(x_i-\langle x_i \rangle )^2}{\sigma_i^2}\thinspace,
\label{eq:chig}\\
\mathrm{Poisson}\quad\chi^2&=& 
\sum_{i=1}^b\left( 2 [\langle x_i \rangle - x_i ]+2
  \thinspace x_i \log \frac{x_i}{\langle x_i \rangle
    }\right)\thinspace ,
\label{eq:chi2}
\end{eqnarray}
where $b$ denotes the number of bins, $x_i$ is the number of events in 
bin $i$ and $\langle x_i \rangle $ denotes the expectation value for 
bin $i$. $\sigma_i$ is estimated by $\sqrt{x_i}$. This method is 
numerically straightforward, but it is not easy to include uncertainties 
of the left out parameters and systematical errors. The Gaussian 
$\chi^2$-function allows, however, an approximate treatment of such
uncertainties by Gaussian error propagation:
\begin{eqnarray}
\chi^2&=&\sum_{ij}(x_i-\langle x_i \rangle )\,
V^{-1}_{ij}\,(x_j-\langle x_j \rangle )\thinspace, 
\label{eq:chige}\\
V_{ij}&=&\sigma_i^2 \delta_{ii} + \sum_{\alpha=1}^n 
\frac{\partial \langle x_i \rangle}{\partial 
\lambda_{\alpha}}\,\frac{\partial \langle x_j \rangle}{\partial 
\lambda_{\alpha}}\,\sigma_{\alpha}^2\thinspace .
\nonumber
\end{eqnarray}
Here $\lambda_{\alpha}$ is one out of $n$ (in our case $n=6$) 
parameters and $\sigma_{\alpha}$ denotes the uncertainty of 
$\lambda_{\alpha}$. This method was for example used 
in~\cite{Burguet-Castell:2001ez}. It is, however, important 
to keep in mind that such an approach relies on the validity 
of the linear approximation of the functional dependence of 
$\langle x_i \rangle$ on $\lambda$. The point is that Gaussian 
error propagation relies only on the linear terms of a Taylor 
expansion of $\langle x_i \rangle$ in $\lambda$. The above 
approach may thus be no longer justified as soon as the second 
or higher order terms become important.

Using one of eqs.~\ref{eq:chig}-\ref{eq:chige}, the acceptance 
region $M$ at confidence level $\alpha$ is given by:
\begin{eqnarray}
  \chi^2(\hat\lambda)&\leq
&\chi^2(\lambda)\quad \forall \lambda \thinspace,
\label{eq:bestfit}\\
\Delta\chi^2(\lambda)&:=&
\chi^2(\lambda)- \chi^2(\hat\lambda)\thinspace,\nonumber\\
M&=&\{\lambda:\Delta\chi^2(\lambda)\leq\Delta\chi^2_{\alpha}\}
\thinspace.
\end{eqnarray}
Using the asymptotic properties of maximum likelihood estimators,
$\Delta\chi^2_{\alpha}$ is determined from the condition that the 
area under a $\chi^2$-distribution from zero to 
$\Delta\chi^2_{\alpha}$ equals $\alpha$. The number of degrees of 
freedom for this distribution is given by the number of free 
parameters $r$ in the fit. The error on the confidence level 
$\alpha$ is then given by the size of the $r$-dimensional 
manifold\footnote{Note that $M$ is not necessarily connected and 
  it may have a quite complicated topological structure, 
  especially if there are parameter degeneracies.} $M$.  This
procedure leads to an error interval $I_{\lambda_i}$ for each 
of the $r$ fitted parameters by simply projecting $M$ onto 
$\lambda_i$. We define the infimum and supremum with respect 
to $\lambda_i$, i.e. the lower (upper) bound on $\lambda_i$:
\begin{eqnarray}
  \inf_{\lambda_i}(M)&:=&\inf(\{\lambda_i:\lambda \in M\})\thinspace
  ,\nonumber\\
 \sup_{\lambda_i}(M)&:=&\sup(\{\lambda_i:\lambda \in M\})\thinspace .
\end{eqnarray}
$I_{\lambda_i}$ is then given by:
\begin{equation}
I_{\lambda_i}=\left[\inf_{\lambda_i}(M),\sup_{\lambda_i}(M) \right]\thinspace .
\end{equation}
This method leads for $n=2$ and a Gaussian likelihood to the usual 
error ellipses.

\begin{figure}[t!]
\begin{center}
\includegraphics[width=0.9\textwidth,angle=0]{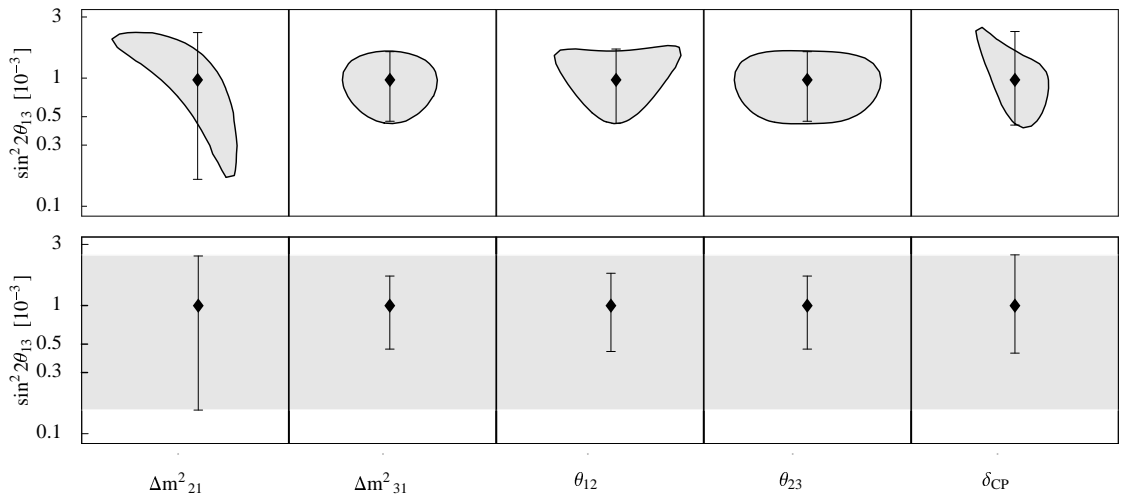}
\end{center}
\caption{Illustration of the error determination for a single  
  parameter (in this example $\sin^2 2\theta_{13}$). In the upper
  row $\sin^2 2\theta_{13}$ is fitted against the remaining 5 
  other parameters (grey regions) and the individual error bars 
  of our method for each parameter are overlayed. These error bars
  result from the projection of the grey regions onto 
  $\sin^2 2\theta_{13}$.  
  All these errors are then combined into a total error for
  $\sin^22\theta_{13}$, given by the lowest and highest endpoints
  of the individual errors. The result is shown as a grey region 
  in the lower row, where the individual errors are overlayed.}
\label{fig:statistik}
\end{figure}

We use a method which we call the 
{ {G}eneralized {C}ovariance {M}atrix {M}ethod}
({\sc gcmm}), where $\chi^2$ is calculated according to eq.~\ref{eq:chi2}
with $n=6$ parameters, namely $\lambda=(\dm{21},\dm{31},$
$\theta_{12},\theta_{13},\theta_{23}, \deltacp)$. The computational effort 
required to calculate the full six-dimensional manifold $M$ is 
enormous\footnote{Even with our approximation below this analysis 
  required to compute $\geq 10^8$ event rate spectra for each 
  channel, which amounts roughly to $1,000\thinspace\mathrm{h}$ 
  CPU time on a $1\thinspace\mathrm{GHz}$ Pentium~III processor}. 
To reduce this effort we adopted a scheme, which takes all 
2-parameter-correlations into account. We calculate therefore all 
two dimensional slices $S_{ij}$ of $M$ through $\hat\lambda$ which 
are parallel to the coordinate planes in order to estimate the 
errors for all six parameters:
\begin{eqnarray}
P_{ij}&=&\{\lambda:\lambda_{k}\equiv \hat\lambda_k;k 
\not = i,j\}\thinspace ,\nonumber\\
S_{ij}&=&\{\lambda\in P_{ij}:\Delta\chi^2(\lambda)\leq
\Delta\chi^2_{\alpha}\}\thinspace .
\end{eqnarray}
These slices can be seen in the upper row of 
fig.~\ref{fig:statistik}, where the method is graphically explained.
There are five such slices for one parameter and the largest 
extension of the projection of these five slices down to the 
parameter of interest is the corresponding error of this 
parameter. The error interval $I_{\lambda_i}$ of a parameter
$\lambda_i$ is thus given by
\begin{eqnarray}
I_{\lambda_i}&=&\bigcup_{k\not = i}\left[\inf_{\lambda_i}(S_{ik}),\sup_{\lambda_i}(S_{ik}) \right]\thinspace ,
\end{eqnarray}
which corresponds to the grey band in the lower row of 
fig.~\ref{fig:statistik}. We use for $\Delta\chi^2_{\alpha}$ the value 
$9.2$ which corresponds to 99\%~C.L. for two degrees of freedom, 
and to 83.8\%~C.L. for six degrees of freedom. The big advantage of 
the {\sc gcmm} method is that no choice about ``most influential''
parameters has to be made. Instead, the method automatically 
finds the most influential parameters, like in the example of 
fig.~\ref{fig:statistik}, where the biggest error for $\theta_{13}$ 
surprisingly comes from $\dm{21}$. Previous neutrino factory studies
used so far only one of the $S_{ij}$ to estimate the accuracy to
which parameters can be extracted. We will show in this work
that this can be a poor approximation, since some of the parameters, 
like $\theta_{13}$ and $\dm{21}$, are strongly correlated in a
previously overlooked way. We will see that the sensitivity to 
$\szt{13}$ can be reduced in this way up to one order of 
magnitude. Another advantage of the {\sc gcmm} method is, that it
can cope with uncertainties of all parameters even if the parameter 
dependence is highly non-linear, as it is often the case in 
neutrino oscillation studies. The {\sc gcmm} method is thus
in a number of ways better than a Gaussian error propagation.
The {\sc gcmm} is, however, still only an approximation, and 
full six parameter fits will lead to corrections. We compared 
the {\sc gcmm} method to three parameter fits and found that 
it works quite good, but the difference can in the worst case 
amount to a factor of two. 

A final aspect of the analysis concerns the inclusion of parameters
from other experiments. This leads to an understanding of how the
parameter improvements of other experiments affect the analysis.
An important case is $\dm{21}$, which can be measured by 
KamLAND~\cite{BARGER} in the LMA-MSW case with a precision which
is much better than what can be obtained at a neutrino factory,
as will be discussed in detail in section~\ref{sec:SUBSUBLEADING}. 
It is therefore important to perform an analysis, where such 
external information can be included and where its impact can 
be assessed. Usually we extract in our analysis parameters by
finding their best fit value and by determining the errors with
the fitting procedure to our ``data''. In order to include external 
knowledge of some parameter (e.g. on $\dm{21}$), we simply restrict 
the range of variation of the is parameter\footnote{In a strict 
  frequentist sense this leads to the loss of coverage. It should 
  be understood as a uniform prior in the Bayesian sense on this 
  parameter.} 
in the fits to the error interval at confidence level $\alpha$.

For the CP-phase there will, however, exist no information in 
advance. Since $\deltacp$ determines which correlations are 
important, the errors on specific parameters can significantly 
depend on the value of $\deltacp$. In such cases, the analysis 
is done for all possible values of $\deltacp$ and the maximal error 
which appears is taken as the final error. We call this procedure 
``$\deltacp$ unknown''. It is equivalent to integrating out a 
nuisance parameter with uniform prior in a Bayesian analysis.

%%%%%%%%%%%%%%%%%%%%%%%%%%%%%%%%%%%%%%%%%%%%%%%%%%%%%%%%%%%%%
\section{Results}\label{sec:results}
%%%%%%%%%%%%%%%%%%%%%%%%%%%%%%%%%%%%%%%%%%%%%%%%%%%%%%%%%%%%%

The results obtained from the numerical study are presented following the
outline given in sec.~\ref{sec:classification}. First, the leading parameters
$\theta_{23}$ and $\dm{31}$ are discussed in sec.~\ref{sec:LEADING}.  Then,
measurements of $\theta_{13}$, matter effects and the sign of $\dm{31}$ are
studied in sec.~\ref{sec:SUBLEADING}. Finally, CP-violating effects and
measurements of $\dm{21}$ and $\theta_{12}$ are presented in
sec.~\ref{sec:SUBSUBLEADING}.  The reader can select certain topics without
missing information which is necessary for the understanding since the
contents of these sections widely do not depend on each other.

%%%%%%%%%%%%%%%%%%%%%%%%%%%%%%%%%%%%%%%%%%%%%%%%%%%%%%%%%%%%%%%%%%%%%
\subsection{\boldmath Leading parameters $\theta_{23}$ and $\dm{31}$}
\label{sec:LEADING}
%%%%%%%%%%%%%%%%%%%%%%%%%%%%%%%%%%%%%%%%%%%%%%%%%%%%%%%%%%%%%%%%%%%%%

First, we will discuss the statistical sensitivity to the leading parameters
$\theta_{23}$ and $\dm{31}$. Neutrino factory experiments will allow high
precision measurements of these two parameters, which are at present only
roughly determined. The information on these parameters dominantly stems from
disappearance channel measurements. Thus, the dependence on the value of
$\theta_{13}$ is marginal, especially when $\theta_{13}$ is small.  It will be
demonstrated that the influence from uncertainties on $\dm{21}$ and
$\theta_{21}$ can however be substantial. This has the consequence that input
from other experiments (like KamLAND) can be helpful to limit the errors of
the affected measurements.

%%%%%%%%%%%%%%%%%%%%%%%%%%%%%%%%%%%%%%%%%%%%%%%%%%%%%%%%%%%%%%%%%%%%%
\subsubsection*{\boldmath Optimization of baseline $L$ and muon energy $E_\mu$}
%%%%%%%%%%%%%%%%%%%%%%%%%%%%%%%%%%%%%%%%%%%%%%%%%%%%%%%%%%%%%%%%%%%%%

\begin{figure}[ht!]
\begin{center}
  \includegraphics[width=0.9\textwidth,angle=0]{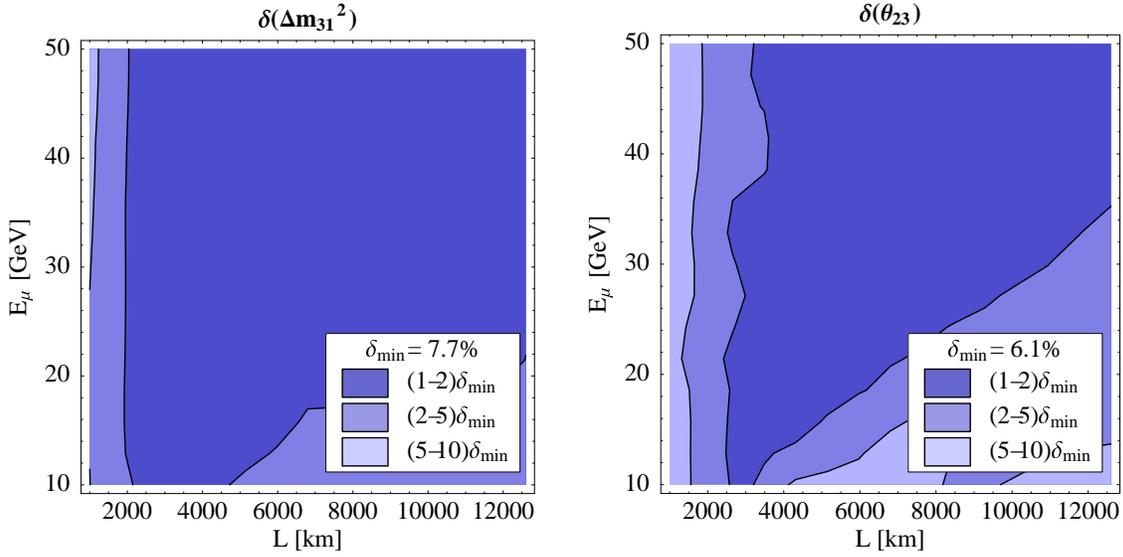}
\end{center}
\caption{Statistical (relative) error of $\dm{31}$ (left plot) and 
  $\sin^2 2\theta_{23}$ (right plot)
  as function of the baseline $L$ and the muon energy $E_\mu$ for
  $\dm{31}=3.5\cdot 10^{-3}\eV^2$, $\sin^2 2\theta_{23}=1$ and $N_\mu \,
  m_\mathrm{kt} = 2\cdot 10^{21} \,\mathrm{kt}\,\mathrm{year}$. Dark shading
  indicates the preferred regions. For $\delta(\dm{31})$, the best sensitivity
  reached is $\delta_\mathrm{min} = 7.7\%$.  For $\delta(\dm{31})$ it is
  $6.1\%$. The contour lines indicate factors of 2, 5 and 10 accuracy loss
  compared to the best values. The \Psl~parameters $\sin^2 2\theta_{13}$,
  $\dm{21}$ and $\deltacp$ do not play a crucial role and are assumed as
  unknown (see sec.~\ref{sec:statistics}).}
\label{fig:LEoptLeading}
\end{figure}
Fig.~\ref{fig:LEoptLeading} shows the expected statistical error on the
quantities $\dm{31}$ and $\sin^2 2\theta_{23}$ in dependence of the neutrino
factory parameters, baseline $L$ and muon energy $E_\mu$ at $N_\mu \,
m_\mathrm{kt} = 2\cdot 10^{21} \,\mathrm{kt}\,\mathrm{year}$.  For beam
energies above $20\GeV$ and baselines above $2000\,$km, the statistical error
on $\dm{31}$ takes values between roughly $8\%$ and $16\%$. For not too small
muon energies, the important constraint is the baseline limit. The above
mentioned $2000\,$km define the limit where the accuracy loss relative to the
best value is less then a factor of two.  This baseline limit linearly depends
on the inverse of $\dm{31}$, which was chosen to be $3.5\cdot 10^{-3}\eV^2$ in
the above case. For $\dm{31} = 1.0 \cdot 10^{-3}\eV^2$ the limit shifts to
approximately $5000\,$km above which the statistical error is less than 20\%.
For $\dm{31} = 6.5 \cdot 10^{-3}\eV^2$ the limit is well below $1000\,$km and
the statistical error does not exceed 14\%.

The statistical error in measurements of the mixing angle $\theta_{23}$ (right
plot of fig.~\ref{fig:LEoptLeading}) is, with minimally 6\%, at the same level
of the one on $\dm{31}$.  Also here, a baseline limit can be given under which
the statistical error is larger than twice the best achievable value. For
$\dm{31} = 3.5\cdot 10^{-3}\eV^2 (1.0\cdot 10^{-3}\eV^2)$ this limit is
roughly $3000\,$km ($5000\,$km). The figure indicates that there is a favored
value of $L/E_\mu$, which can easily be understood using analytic
considerations: Information on $\theta_{23}$ is dominantly extracted from the
total rates observed in the disappearance channel $P(\nu_\mu \rightarrow
\nu_\mu)$ which is is not significantly affected by matter effects.  Assuming
a discrete neutrino energy corresponding to the average neutrino energy in the
disappearance channel ($\left<E_\nu\right> = 16/21 E_\mu$), the number of
observed muon neutrino events is approximately given by $N = (1-\sin^2
2\theta_{23}\sin^2 \hat\Delta_{31}) \Phi_{\nu_\mu}$, where the factor
$\Phi_{\nu_\mu}$ includes flux, cross-sections and detector mass.  For $N_\mu
\, m_\mathrm{kt} = 2\cdot 10^{21} \,\mathrm{kt}\,\mathrm{year}$, we find that
$\Phi_{\nu_\mu} = 6.9 10^6 (L/\mathrm{km})^{-2}(16/21 E_\mu/\GeV^2)^3$. In the
Gaussian limit, which is a good approximation here, the relative statistical
error on the quantity $\sin^2 2\theta_{23}$ is given by
\begin{equation}
\label{eq:th23err}
\left|\frac{\Delta \sin^2 2\theta_{23}}{\sin^2 2\theta_{23}}\right| = 
\frac{1}{\sqrt{\Phi_{\nu_\mu}}}
\frac{\sqrt{1-\sin^2 2\theta_{23} \sin^2 
\hat\Delta_{31}}}{\sin^2 2\theta_{23} \sin^2 \hat\Delta_{31}}
= f(\frac{L}{E_\mu}) E_\mu^{-1/2} ~.
\end{equation}
Thus, the relative error on $\sin^2 2\theta_{23}$ is approximately a function
of $L/E_\mu$ with a slight modification from the factor $E_\mu^{-1/2}$. For
muon energies between $10\GeV$ and $50\GeV$ this modification maximally gives
a correction of $(50/10)^{-1/2} \approx 0.45$ which in general reduces the
error for higher muon energies. The corresponding plot in
fig.~\ref{fig:LEoptLeading} clearly shows the contour lines of constant
$L/E_\mu$. Using eq.~\ref{eq:th23err} with the parameters corresponding to
fig.~\ref{fig:LEoptLeading} yields a relative error for $\sin^2 2\theta_{23}$
at the level 1/1000. For the quantity $\theta_{23}$ this translates to a
relative error at the level of some per cent, which is in good agreement with
the results obtained in the full numerical analysis.

Summarily, the leading parameters do not give very strong recommendations for
the selection of the baseline and muon energy. Beam energies between $30\GeV$
and $50\GeV$ are recommended. To achieve a sufficiently developed oscillation,
the baseline must be large enough. For the central value of $\dm{31}$ proposed
by the Super-Kamiokande experiment, a baseline of $3000\,$km or more would be
suitable. Should it turn out in the future, that $\dm{31}$ is at the lower
edge of the presently favored region, one should be aware that this baseline
limit shifts to higher values roughly inverse proportionally to $\dm{31}$.
The exact values of the \Psl~parameters $\theta_{13}$ and $\deltacp$ do not
play a role in this discussion since the results presented here depend only
weakly on them. But $\dm{21}$ being correlated with $\dm{31}$ can have some
impact on the errors made in measurements of $\dm{31}$. This point will be
discussed in detail later.

%%%%%%%%%%%%%%%%%%%%%%%%%%%%%%%%%%%%%%%%%%%%%%%%%%%%%%%%%%%%%%%%%%%%%
\subsubsection*{\boldmath Dependance on flux and detector 
mass $N_\mu \, m_\mathrm{kt}$}
%%%%%%%%%%%%%%%%%%%%%%%%%%%%%%%%%%%%%%%%%%%%%%%%%%%%%%%%%%%%%%%%%%%%%

\begin{figure}[ht!]
\begin{center}
  \includegraphics[width=0.3\textwidth,angle=-90]{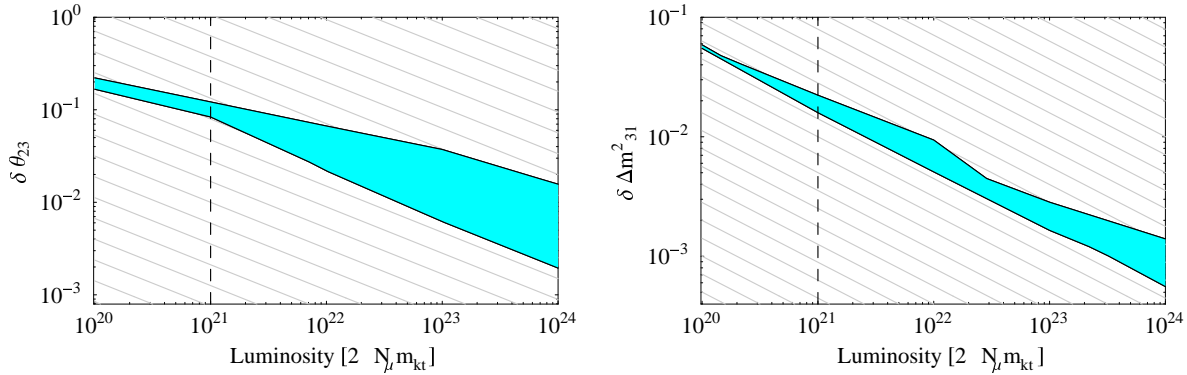}
\end{center}
\caption{Statistical error of $\theta_{23}$ (left plot) 
  and $\dm{31}$ (right plot) as function of $N_\mu \, m_\mathrm{kt}$, the
  number of stored muons per year times the mass of the neutrino detector in
  kilotons.  The shaded bands indicate the range which is covered by
  variations of the \Psl parameters $\theta_{13}$ and $\dm{21}$ ($0 \le \sin^2
  2\theta_{13} \le 10^{-1}$, $10^{-5} \le \dm{21} \le 10^{-4}$). The
  calculation of $\delta(\theta_{23})$ was performed with $50\GeV$ muon energy
  and a baseline of $7000\,$km ($\delta\theta_{23}$) respectively $3000\,$km
  ($\delta\dm{31}$).}
\label{fig:fluxTh23}
\end{figure}

The influence of $N_\mu \, m_\mathrm{kt}$, the number of stored muons per year
times the mass of the neutrino detector in kilotons, is shown in
fig.~\ref{fig:fluxTh23}. The decrease of the statistical error with increasing
$N_\mu \, m_\mathrm{kt}$ roughly follows the prediction from Gaussian
statistics ($1/\sqrt{N_\mu \, m_\mathrm{kt}}$), which is indicated by the
parallel lines in the background of the figure. The spread of the shaded band
is generated by variations of the \Psl~parameters. Their influence increases
with higher luminosities where the statistical errors are small.

%%%%%%%%%%%%%%%%%%%%%%%%%%%%%%%%%%%%%%%%%%%%%%%%%%%%%%%%%%%%%%%%%%%%%
\subsubsection*{\boldmath Correlation between $\dm{31}$ and $\dm{21}$}
%%%%%%%%%%%%%%%%%%%%%%%%%%%%%%%%%%%%%%%%%%%%%%%%%%%%%%%%%%%%%%%%%%%%%

\begin{figure}[ht!]
\begin{center}
  \includegraphics[width=0.6\textwidth,angle=0]{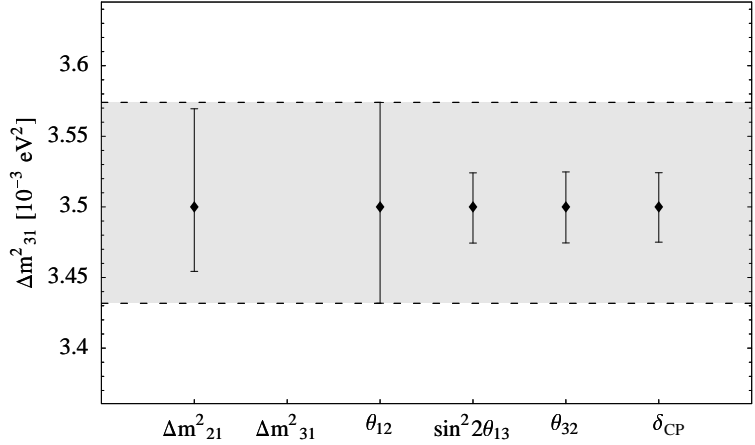}
\end{center}
\caption{Error bars of all two parameter fits which include $\dm{31}$. 
  The total error 
  (grey shaded region) is dominated by the contributions from the fits against
  $\dm{21}$ and $\theta_{12}$. Fitting $\dm{31}$ only against $\theta_{23}$
  can result in a considerable underestimation of $\delta(\dm{31})$. The
  calculation was performed with $L=8000\,$km, $E_\mu = 50\GeV$, $\theta_{12}
  = \pi/4$, $\sin^2 2\theta_{13} = 10^{-3}$, $\dm{21} = 10^{-4}\eV^2$ and
  $N_\mu \, m_\mathrm{kt} = 2\cdot 10^{21} \,\mathrm{kt}\,\mathrm{year}$.  }
\label{fig:errorbarsDm31}
\end{figure}

\begin{figure}[ht!]
\begin{center}
\includegraphics[width=0.3\textwidth,angle=-90]{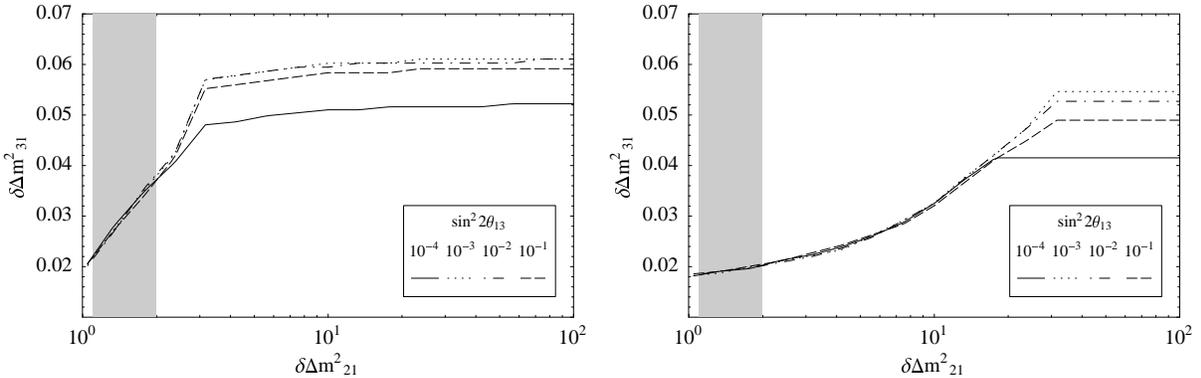}
\end{center}
\caption{Statistical error on 
  $\dm{31}$ as function of $\delta(\dm{21})$, the error on $\dm{21}$ as
  provided from external measurements (like the KamLAND experiment).  The
  shaded band indicates the expected values of $\delta(\dm{21})$ from
  measurements performed at the KamLAND experiment \cite{BARGER}. No external
  input corresponds to the right edge of the plot ($\delta(\dm{21})=10^{2}$).
  The calculation was done for $\dm{21} = 10^{-4}\eV^2$ (left panel)
  respectively for $\dm{21} = 10^{-5}\eV^2$ (right panel) and
  $\theta_{12}=\pi/4$. The different line styles indicate different values of
  $\sin^2 2\theta_{13}$ ($10^{-4}$, $10^{-3}$, $10^{-2}$, $10^{-1}$).}
\label{fig:dm21leading4_8000}
\end{figure}

It is important to note that, for specific selections of $L$ and $E_\mu$, the
statistical error on $\dm{31}$ is dominated by its correlation with $\dm{21}$.
This fact is displayed in fig.~\ref{fig:errorbarsDm31}, which shows the
results of two parameter fits of $\dm{31}$ with all other oscillation
parameters. The plot shows that the dominating contribution to the statistical
error of $\dm{31}$ stems from the uncertainties of $\dm{21}$ and
$\theta_{12}$.  This suggests that in such cases, external information on
$\dm{21}$ and $\theta_{12}$ from other experiments can considerably improve
the precision of measurements of $\dm{31}$.
With fig.~\ref{fig:dm21leading4_8000}, this point is investigated in more
detail: There, the resulting statistical error on $\dm{31}$ is given as
function of $\delta(\dm{21})$, the error on $\dm{21}$ as provided from
external measurements. The shaded band indicates the expected values of
$\delta(\dm{21})$ from measurements performed at the KamLAND
experiment~\cite{BARGER}.  The plots show that KamLAND input can improve the
statistical error on $\dm{31}$ up to a factor of three.  This, however, is
only valid for specific baselines. At smaller baselines this correlation is
not very pronounced.

%%%%%%%%%%%%%%%%%%%%%%%%%%%%%%%%%%%%%%%%%%%%%%%%%%%%%%%%%%%%%%%%%%%%%
\subsection{\boldmath Sub-leading parameters $\theta_{13}$ and 
$\mathrm{sgn} \dm{31}$}
\label{sec:SUBLEADING}
%%%%%%%%%%%%%%%%%%%%%%%%%%%%%%%%%%%%%%%%%%%%%%%%%%%%%%%%%%%%%%%%%%%%%
%
Following the outline given in section~\ref{sec:classification}, we continue
this study with a discussion of the parameter $\theta_{13}$. Information on
$\theta_{13}$ is mainly obtained from the appearance channel probabilities
$P(\nu_e \rightarrow \nu_\mu)$ and $P(\bar{\nu}_e \rightarrow \bar{\nu}_\mu)$,
which contain terms proportional to $\sin 2\theta_{13}$ and $\sin^2
2\theta_{13}$.  For not too small values of $\sin^2 2\theta_{13}$, the results
presented in this section are roughly independent from the \Pssl~parameters
$\dm{21}$, $\theta_{12}$ and $\deltacp$. In this case, a neutrino factory
experiment will be able to measure $\theta_{13}$ with some precision.
However, for small values of $\sin^2 2\theta_{13}$ close to the sensitivity
reach, the parameters $\dm{21}$, $\theta_{12}$ and $\deltacp$ can have
considerable impact on measurements of $\theta_{13}$. This is an example where
the step-wise analysis of such experiments is no longer valid.

%%%%%%%%%%%%%%%%%%%%%%%%%%%%%%%%%%%%%%%%%%%%%%%%%%%%%%%%%%%%%%%%%%%%%
\subsubsection*{\boldmath Optimization of baseline $L$ and muon energy $E_\mu$}
%%%%%%%%%%%%%%%%%%%%%%%%%%%%%%%%%%%%%%%%%%%%%%%%%%%%%%%%%%%%%%%%%%%%%

\begin{figure}[ht!]
\begin{center}
  \includegraphics[width=\textwidth,angle=0]{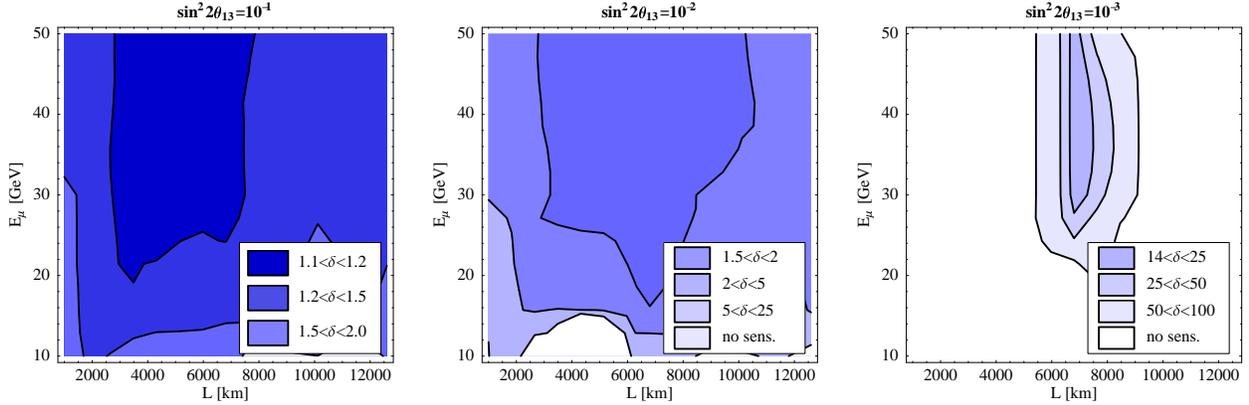}
\end{center}
\caption{Statistical error of the quantity $\sin^2 2\theta_{13}$
  as function of the baseline $L$ and the muon energy $E_\mu$ for
  $\dm{31}=3.5\cdot 10^{-3}\eV^2$, $\sin^2 2\theta_{23}=1$, $N_\mu \,
  m_\mathrm{kt} = 2\cdot 10^{21} \,\mathrm{kt}\,\mathrm{year}$ and three
  values of $\sin^2 2\theta_{13}$ ($10^{-1}$, $10^{-2}$, $10^{-3}$).  Dark
  shading indicates the preferred regions. The parameters $\dm{21}$ and
  $\deltacp$ play a role mainly for small values of $\theta_{13}$. Here, they
  are assumed as unknown (see sec.~\ref{sec:statistics}).}
\label{fig:LEoptSubleading}
\end{figure}
\begin{table}[ht!]
\begin{center}
  \begin{tabular}{|c||c|c|c|c|}
\hline
$\sin^2 2\theta_{13}$ & $10^{-1}$ &  $10^{-2}$ &  $10^{-3}$ &  $10^{-4}$ \\ 
\hline \hline
$\dm{31}=6.0\cdot 10^{-3} \eV^2$ & 1.1 & 1.2 & 2.1 & 10 \\ \hline
$\dm{31}=3.5\cdot 10^{-3} \eV^2$ & 1.1 & 1.5 & 14 & -- \\ \hline
$\dm{31}=1.0\cdot 10^{-3} \eV^2$ & 3.3 & 180 & -- & -- \\ \hline
\end{tabular}
\caption{\label{tab:th13sens}Statistical errors $\delta$ made 
  in measurements of $\sin^2 2\theta_{13}$ 
  listed for several values of $\dm{31}$. The quantity $\delta$, whose values
  are given in the table is the same as in fig.~\ref{fig:LEoptSubleading}. A
  detailed description of $\delta$ is given in the text.}
\end{center}
\end{table}
The statistical error of $\sin^2 2\theta_{13}$ in dependence of the neutrino
factory parameters baseline $L$ and beam energy $E_\mu$ is displayed in
fig.~\ref{fig:LEoptSubleading}.  The plotted quantity is
\begin{equation}
\delta := \mathrm{max}\{(z_\mathrm{min}/z_0)^{-1}, z_\mathrm{max}/z_0\}~,
\end{equation}
where $z=\sin^2 2\theta_{13}$ and the subscripts correspond to the central
value (0), the minimal value (min) and the maximal value (max) which are
compatible with the simulated experimental data.  For values of $\delta$ close
to 1, $\delta-1$ is approximately equal to the relative error. $\delta = 10
(100)$ indicates one (two) order(s) of magnitude uncertainty on the measured
quantity.  Since the error strongly depends on the value $\sin^2
2\theta_{13}$, three plots for the values $\sin^2 2\theta_{13}=(10^{-1},
10^{-2}, 10^{-3})$ are given. For $\sin^2 2\theta_{13}=10^{-1}$ (left plot),
the sensitivity is good and reaches values between 10\% and 20\% relative
error at baselines between $3000\,$km and $8000\,$km and with beam energies
above $20\GeV$. For $\sin^2 2\theta_{13}=10^{-2}$ (middle plot), still errors
down to 50\% are reachable. Close to the sensitivity reach $\sin^2
2\theta_{13}=10^{-3}$ (right plot), the situation is interesting. There, the
information is still sufficient to determine $\sin^2 2\theta_{13}$ with an
error corresponding to roughly one order of magnitude. But now, close to the
sensitivity limit, long baselines between $7000\,$km and $8000\,$km are
strongly preferable. The reason for this is, that for small values of $\sin^2
2\theta_{13}$, the correlations of $\theta_{13}$ with $\dm{21}$ and $\deltacp$
are important. These correlations tend to have less influence at large
baselines. Since previous studies concerning the optimization of the
baseline~\cite{CERVERA} did not include these correlations, they missed this
point and came to differing recommendations for the baseline.  Note, that
close to the sensitivity limit, the result depends strongly on the precise
value of the CP phase $\deltacp$. For the calculation which is presented here,
we assume that $\deltacp$ is completely unknown: The statistical errors are
computed for several values of $\deltacp$\footnote{uniformly covering an
  interval of $2\pi$} and the maximal error of this set is considered as the
final result.

In table \ref{tab:th13sens} the statistical errors made in measurements of
$\sin^2 2\theta_{13}$ are listed for several values of $\dm{31}$. The value of
$\dm{31}$ changes the obtained results significantly. Higher values of
$\dm{31}$ increase the precision, whereas lower values make it more difficult
to measure $\theta_{13}$.

%%%%%%%%%%%%%%%%%%%%%%%%%%%%%%%%%%%%%%%%%%%%%%%%%%%%%%%%%%%%%%%%%%%%%
\subsubsection*{\boldmath Correlation between $\theta_{13}$ and the
  \Pssl~parameters $\dm{21}$ and $\deltacp$} 
%%%%%%%%%%%%%%%%%%%%%%%%%%%%%%%%%%%%%%%%%%%%%%%%%%%%%%%%%%%%%%%%%%%%%

In connection with measurements of $\deltacp$ it has been recognized that the
correlation of the CP phase $\deltacp$ with the mixing angle $\theta_{13}$
could possibly hide the effect of CP violation~\cite{Burguet-Castell:2001ez}.
The implications of this correlation on the expected precision of measurements
of of $\theta_{13}$ have not been discussed.  Already in the baseline
discussion above, it was shown that in case of small values of $\theta_{13}$,
correlations with $\deltacp$ and $\dm{21}$ can drastically influence the
statistical errors expected in measurements of $\sin^2 2\theta_{13}$.
\begin{figure}[ht!]
\begin{center}
\includegraphics[height=\textwidth,angle=-90]{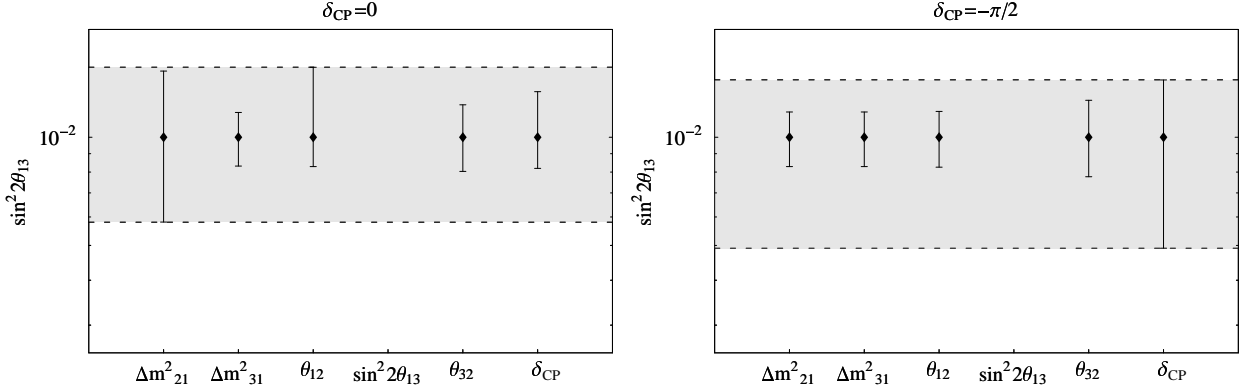}
\end{center}
\caption{Error bars of all two parameter fits which 
  include $\sin^2 2\theta_{13}$ for $\deltacp = 0$ (left plot)
  and $\deltacp = \pi/2$ (right plot). The total error (grey shaded region) is
  dominated by the contributions from the fits against $\dm{21}$ (left plot)
  or $\deltacp$ (right plot). Fitting $\sin^2 2\theta_{13}$ only together with
  $\dm{31}$ can result in a considerable underestimation of the statistical
  error. The calculation was performed with $L=2100\,$km, $E_\mu = 50\GeV$,
  $\theta_{12} = \pi/4$, $\sin^2 2\theta_{13} = 10^{-2}$, $\dm{21} =
  10^{-4}\eV^2$ and $N_\mu \, m_\mathrm{kt} = 2\cdot 10^{21}
  \,\mathrm{kt}\,\mathrm{year}$.}
\label{fig:errorbarsTh13}
\end{figure}
To illustrate this, fig.~\ref{fig:errorbarsTh13} shows the results of two
parameter fits of $\sin^22\theta_{13}$ with all other oscillation parameters
for two different values of the CP phase $\deltacp$. The plots demonstrate
that the dominating contribution to the statistical error of $\sin^2
2\theta_{13}$ can be caused by correlations with the parameters $\dm{21}$ or
$\deltacp$. In the case $\deltacp = 0$ (left plot), the main error stems from
the correlation with $\dm{21}$ but for $\deltacp = \pi/2$ (right plot), it is
the CP phase $\deltacp$ which gives the dominating contribution to the total
error. The correlation with $\dm{21}$ severely affects the sensitivity to
$\sin^2 2\theta_{13}$ which was completely overlooked up to now.

The influence of the \Pssl~parameters $\dm{21}$ and $\deltacp$ for the
measurement of $\theta_{13}$ can qualitatively be understood using analytic
considerations. For small values of $\theta_{13}$ at the sensitivity limit,
all four terms of equation~\ref{eq:PROBMATTER} are equally important. The
first, leading term has always a positive sign. The second term $\propto
\alpha\,\sin \deltacp \,\sin^2 2\theta_{13}$ has a sign which depends on
$\deltacp$ and whether neutrinos or anti-neutrinos are considered. Since we
always use neutrinos and anti-neutrinos, this term can have both signs. The
sign of the third term $\propto \alpha\,\cos \deltacp \,\sin^2 2\theta_{13}$
is determined by the value of $\deltacp$. The sign of the fourth term $\propto
\alpha^2$ is always positive. If $\cos\deltacp<0$ an increase (decrease) in
$\theta_{13}$ can be compensated by an increase (decrease) in $\dm{21}$. In
this case the main problem arises from the correlation of $\theta_{13}$ and
$\deltacp$, since all terms containing $\deltacp$ are proportional to $\dm{21}
\sin^2 2\theta_{13}$ and this product can not be determined very well. Also
the event rates are considerably smaller in this case which simply leads to a
loss of statistical significance especially close to the sensitivity limit. If
$\cos \deltacp>0$ an increase (decrease) in $\theta_{13}$ can be compensated
by a decrease (increase) in $\dm{21}$, therefore the product $\dm{21} \sin^2 2
\theta_{13}$ can be determined quite accurately. In this case where the
correlation of $\sin^2 2\theta_{13}$ with $\dm{21}$ causes the problems,
external information on $\dm{21}$ (e.g. from KamLAND) would help to improve
the results.

It is also possible to understand, why large baselines around $7000\,$km are
recommended in case of low values of $\theta_{13}$ and large values of
$\dm{21}$: The relative size of the terms containing $\deltacp$ decreases with
growing baseline. Thus in principle larger baselines should perform better.
For too long baselines, the $1/L^2$-decrease of event rates starts to worsen
the result again.  The numerical calculation shows that baselines around
$7000\,\mathrm{km}$ perform by far best.  Shorter baselines can profit
considerably from external input on $\dm{21}$ if (and only if) $\cos\deltacp >
0$.

It is of great importance to recognize that for small values of $\theta_{13}$
and large values of $\dm{21}$ the contributions to the total error of
measurements of $\theta_{13}$, originating from the \Pssl~sector, are
substantial and not only minor corrections.  With decreasing $\dm{21}$, these
contributions get smaller and in the limit $\dm{21}=0$ there is no influence
left from the parameters $\dm{21}$, $\theta_{12}$ and $\deltacp$. The same is
true for small values of $\theta_{12}$ (SMA solution).

%%%%%%%%%%%%%%%%%%%%%%%%%%%%%%%%%%%%%%%%%%%%%%%%%%%%%%%%%%%%%%%%%%%%%
\subsubsection*{\boldmath Sensitivity reach for measurements of $\theta_{13}$}
%%%%%%%%%%%%%%%%%%%%%%%%%%%%%%%%%%%%%%%%%%%%%%%%%%%%%%%%%%%%%%%%%%%%%

\begin{figure}[ht!]
\begin{center}
  \includegraphics[width=7cm,angle=-90]{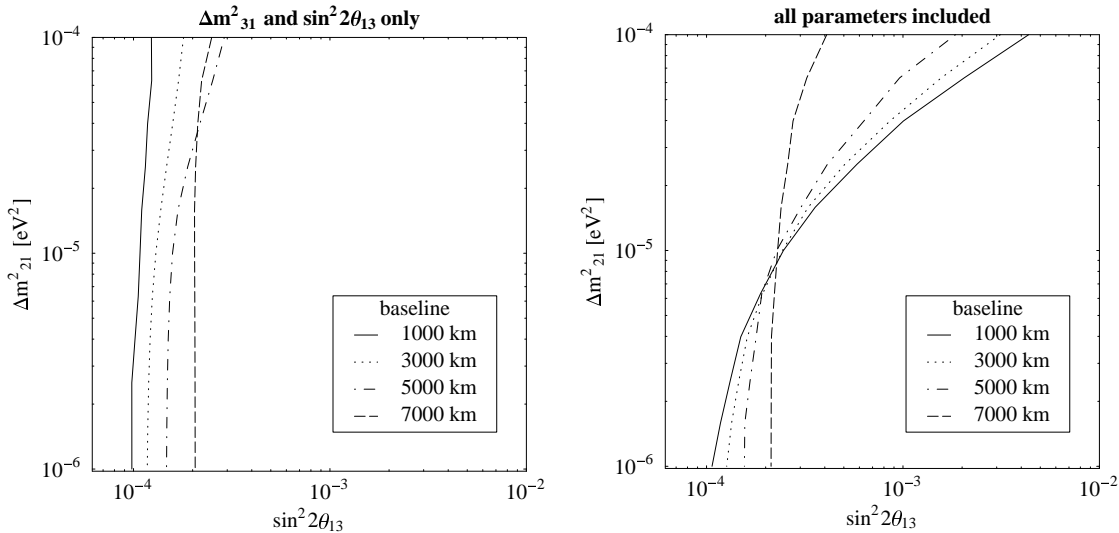}
\end{center}
\caption{Sensitivity reach for measurements 
  of $\sin^2 2\theta_{13}$. The area to the left
  of the lines indicates the parameter range where measurements are compatible
  with $\sin^2 2\theta_{13} = 0$ at 99\%~C.L.\thinspace. The calculation was
  performed with a beam energy of $50\GeV$. The different line types are for
  different baseline as explained in the legend. For comparison to older
  studies, the left panel displays the result obtained from a two parameter
  fit of only $\sin^2 2\theta_{13}$ and $\dm{31}$. The right panel displays
  our new result with all parameters taken into account.}
\label{fig:th13exc3000}
\end{figure}
We define the ``sensitivity reach'' for measurements of the mixing angle
$\theta_{13}$ as the maximal value of $\sin^2 2\theta_{13}$ at which the
experimental data is compatible with the hypothesis $\sin^2 2\theta_{13} = 0$
at 99\%~C.L.\,.  Measurements of nonzero values for $\theta_{13}$ are only
possible above the sensitivity reach. The sensitivity reach as function of
$\dm{21}$ is given by the right hand plot in fig.~\ref{fig:th13exc3000}. The
different line types indicate different baselines. For very small $\dm{21}$,
values down to $\sin^2 2\theta_{13} \simeq 10^{-4}$ are reachable and
baselines between $1000\,\mathrm{km}$ and $7000\,\mathrm{km}$ perform very
similar.  The sensitivity limit varies only within a factor two. With
increasing $\dm{21}$, the effects from the parameters $\dm{21}$ and $\deltacp$
become stronger and the sensitivity reach deteriorates by more than one order
of magnitude (for short baselines).  The left hand plot shows the results
obtained by a two parameter fit to only $\theta_{13}$ and $\dm{31}$.  Since
there, the correlation with $\dm{21}$ and $\deltacp$ are not taken into
account and $\deltacp$ is fixed to zero, the performance of an experiment is
nearly independent of $\dm{21}$. Comparing the two plots, it is obvious that
with this method, for large values of $\dm{21}$, the performance would be
strongly overestimated, especially for baselines below $7000\,\mathrm{km}$.
This result is important since it suggests to use a longer baseline for
measurements of $\theta_{13}$ than usually proposed.

\begin{figure}[ht!]
\begin{center}
\includegraphics[width=0.6\textwidth,angle=0]{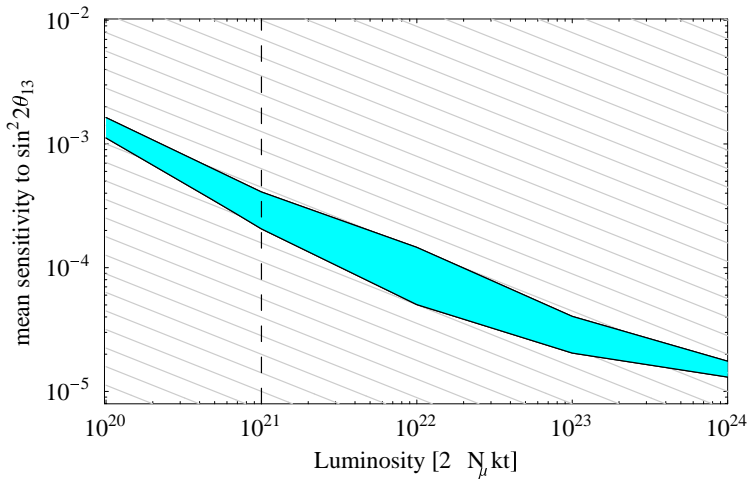}
\end{center}
\caption{Sensitivity reach (see text for detailed definition) 
  for measurements of $\sin^2 2\theta_{13}$ as function of $N_\mu \,
  m_\mathrm{kt}$, the number of stored muons per year times the mass of the
  neutrino detector in kilotons.  The band indicates the range which is
  covered by variations of the \Psl parameters $\dm{21}$ and $\deltacp$
  ($10^{-5} \le \dm{21} \le 10^{-4}$, $-\pi < \deltacp < \pi$).  The
  calculation was performed with a baseline of $7000\,$km and $50\GeV$ muon
  energy. The vertical dashed line indicates a typical neutrino factory
  experiment with a 10~kt iron detector.}
\label{fig:th13flux7000}
\end{figure}

The dependence of the $\theta_{13}$ sensitivity reach on the parameter $N_\mu
\, m_\mathrm{kt}$ is shown in fig.~\ref{fig:th13flux7000}. The scaling
behavior is nearly Gaussian, i.e. it is roughly proportional to $\sqrt{N_\mu
  \, m_\mathrm{kt}}^{-1}$. At high luminosities systematical errors, like
insufficient charge identification capabilities of the detector, will most
probably limit the sensitivity.
%

%%%%%%%%%%%%%%%%%%%%%%%%%%%%%%%%%%%%%%%%%%%%%%%%%%%%%%%%%%%%%%%%%%%%%
\subsubsection*{\boldmath Sensitivity reach for the determination 
of $\mathrm{sgn}\,\dm{31}$}
%%%%%%%%%%%%%%%%%%%%%%%%%%%%%%%%%%%%%%%%%%%%%%%%%%%%%%%%%%%%%%%%%%%%%

The ability to determine the sign of $\dm{31}$ stems from the fact that the
MSW-resonance occurs -- depending on the sign of $\dm{31}$ -- either in the
neutrino or in the anti-neutrino appearance channel. Therefore it is very
important to be sensitive to the energy region around the MSW-resonance.  In
references \cite{FLPR,FHL,Barger:2000cp} it was shown that, in the case this
sensitivity is good, the ability to determine the sign of $\dm{31}$ nearly
coincides with the ability to determine $\theta_{13}$.  This holds in some
approximation also when all parameters are taken into account.  Hence, we do
not study this point in detail but refer to fig.~\ref{fig:th13flux7000}, which
should give also a rough idea about the limit on $\theta_{13}$ above which a
determination of $\mathrm{sgn}\,\dm{31}$ should be possible.  Each term of
eq.~\ref{eq:PROBMATTER} is invariant under a simultaneous change from neutrino
to anti-neutrino and of the sign of $\dm{31}$.  Therefore changing the sign of
$\dm{31}$ is equivalent to interchanging the role of neutrinos and
anti-neutrinos. Assuming symmetric operation of the neutrino factory, the only
difference that remains is due to the cross sections, which are for neutrinos
twice as large as for anti-neutrinos. In case of a negative sign of $\dm{31}$
one thus looses half of the statistics, which can be compensated by doubling
the flux of $\bar\nu_{e}$.

%%%%%%%%%%%%%%%%%%%%%%%%%%%%%%%%%%%%%%%%%%%%%%%%%%%%%%%%%%%%%%%%%%%%%
\subsection{\boldmath Sub-sub-leading parameters $\theta_{12}$, $\dm{21}$ 
and $\deltacp$}
\label{sec:SUBSUBLEADING}
%%%%%%%%%%%%%%%%%%%%%%%%%%%%%%%%%%%%%%%%%%%%%%%%%%%%%%%%%%%%%%%%%%%%%
%
Whether it is possible to measure the parameters $\theta_{12}$, $\dm{21}$ and
$\deltacp$ with a neutrino factory long baseline experiment crucially depends
on the values of $\theta_{13}$, $\theta_{12}$ and $\dm{21}$. To significantly
influence the measured event rates, it is absolutely necessary that the
LMA-MSW solution is the correct description for solar neutrino oscillations.  
In comparison to the results presented in the previous section, which do only
depend on the value of $\theta_{13}$, this is a severe additional condition
which enables or disables the search for the CP-violating phase $\deltacp$.
Assuming now that the LMA-MSW solution is realized, it is important to notice 
that the \Pssl~effects are proportional to $\dm{21}$, which in the LMA-range 
can still vary between roughly $2\cdot 10^{-5}\eV^2$ and 
$2\cdot 10^{-4}\eV^2$ (at 99\% confidence level)\cite{Bahcall:2001hv}. 
This means, that the magnitude of CP-violating effects can also vary by 
one order of magnitude.

In this section we will discuss the following points in detail: First, we will
study the precision of measurements of the parameters $\theta_{12}$ and
$\dm{21}$ in comparison to the KamLAND experiment. We will show, that the
neutrino factory is mainly capable to measure the product $\dm{21} \sin
2\theta_{12}$ but has severe difficulties to constrain $\dm{21}$ and
$\theta_{12}$ separately. KamLAND will provide much better accuracy for these
parameters.  Then, we will focus on the search for CP violation. We will
discuss, for different neutrino factory fluxes and detector masses, the
magnitude of the effects from the CP phase $\deltacp$ in dependence of the
parameters $\theta_{13}$ and $\dm{21}$. Which baselines and beam energies are
best suitable for measurements of the CP phase is also studied in detail. 
Finally, some specific topics like the parameter degeneracy in the 
$\theta_{13}$-$\deltacp$ parameter plane are discussed.

%%%%%%%%%%%%%%%%%%%%%%%%%%%%%%%%%%%%%%%%%%%%%%%%%%%%%%%%%%%%%%%%%%%%%
\subsection*{\boldmath Correlation between $\theta_{12}$ and $\dm{21}$}
\label{sec:nufactvskamland}
\begin{figure}[ht!]
\begin{center}
  \includegraphics[width=0.8\textwidth,angle=0]{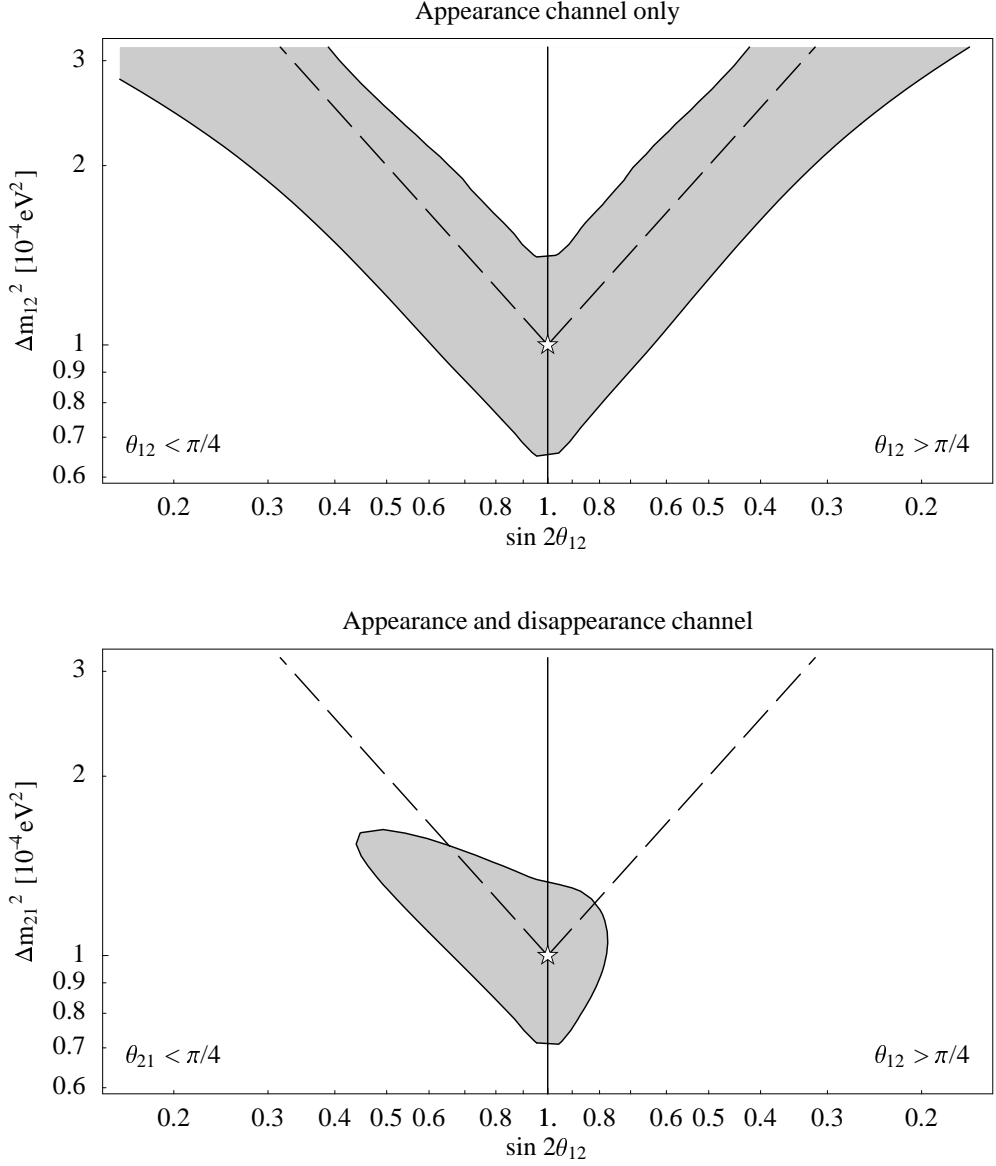}
\end{center}
\caption{$3\sigma$ contour line of two parameter fits of $\theta_{12}$ 
  and $\dm{21}$ to simulated event rates
  from the appearance channel only (upper plot) and appearance channel plus
  disappearance channel (lower plot). Constant values of the product $\dm{21}
  \sin 2\theta_{12}$ are given by the dashed lines.  The calculation was
  performed with $50\,\GeV$ muon energy, $3000\,$km baseline, $N_\mu \,
  m_\mathrm{kt} = 2\cdot 10^{21} ~\,\mathrm{kt}\,\mathrm{year}$,
  $\theta_{12}=\pi/4$, $\dm{21} = 10^{-4}\eV^2$, $\sin^2 2\theta_{13}=
  10^{-3}$ (upper plot) and $\sin^2 2\theta_{13}= 10^{-1}$ (lower plot). }
\label{fig:t12dm21product}
\end{figure}

%%%%%%%%%%%%%%%%%%%%%%%%%%%%%%%%%%%%%%%%%%%%%%%%%%%%%%%%%%%%%%%%%%%%%
%
For $\dm{21} < 10^{-4} \eV^2$, the appearance channel probabilities depend
only on the product $\dm{21} \sin 2\theta_{12}$ (see eq.~\ref{eq:PROBMATTER}).
Thus a strong correlation between $\theta_{12}$ and $\dm{21}$ can be expected.
The upper plot of fig.~\ref{fig:t12dm21product} shows the result of a two
parameter fit
of $\theta_{21}$ and $\dm{21}$ to only the appearance rates.  From the
comparison to the dashed line, which represents the constant value $\dm{21}
\sin 2\theta_{12} = 10^{-4} \eV^2$, it can clearly be seen that indeed the
measurement is only sensitive to the product $\dm{21} \sin 2\theta_{12}$.  For
large values of $\dm{21}$, this correlation can be lifted by inclusion of
disappearance rates (see lower plot of fig.~\ref{fig:t12dm21product}).  The
\Psl~term in the disappearance probability (eq.~\ref{DISPROBVACUUM}) depends
on the product $\dm{21} \cos^2 \theta_{12}$, which helps to lift the
degeneracy between $\dm{21}$ and $\theta_{12}$.  If, however, $\dm{21}$ is
small, the $\alpha^2$-term in the disappearance probability looses in strength
and it gets very difficult to obtain information on $\theta_{12}$ and
$\dm{21}$ separately.

\begin{figure}[ht!]
\begin{center}
\includegraphics[width=0.8\textwidth,angle=0]{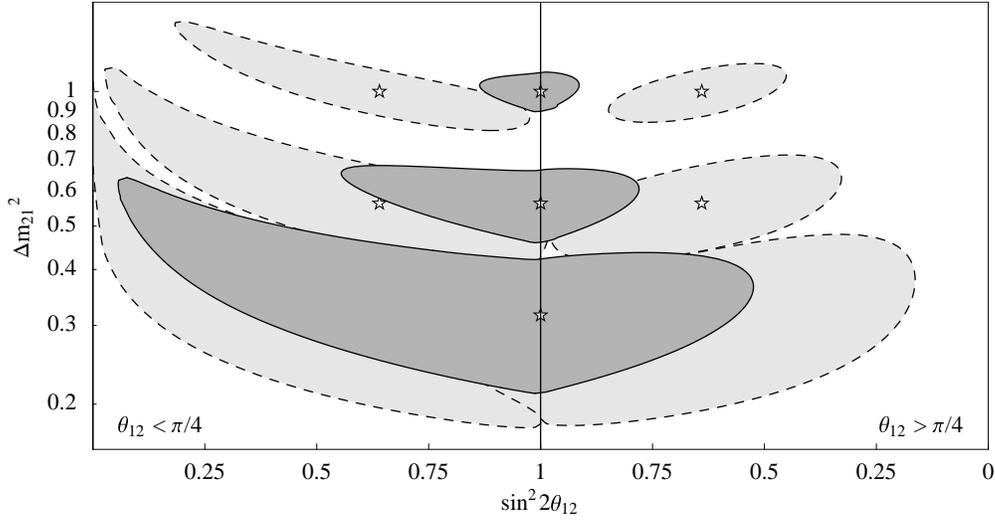}
\end{center}
\caption{Two parameter fits of $\theta_{12}$ and $\dm{21}$ to combined 
simulated event rates of the appearance channel and the disappearance
channel. The calculation was performed with $50\GeV$ muon energy, 
$3000\,$km baseline, $N_\mu \, m_\mathrm{kt} = 2\cdot 10^{22} ~\,
\mathrm{kt}\,\mathrm{year}$ (!!) and $\sin^2 2\theta_{13}= 10^{-1}$.}
\label{fig:t12dm21}
\end{figure}

Hence, the precision of measurements of $\theta_{12}$ and 
$\dm{21}$ is in general not very good. Particularly, the 
performance of a neutrino factory in such measurements is not 
comparable to the KamLAND experiment. Fig.~\ref{fig:t12dm21} 
shows results of two parameter fits of $\theta_{12}$ and $\dm{21}$ 
from the combined rates of the appearance and disappearance 
channels for several values of $\dm{21}$ and $\theta_{12}$.
The simulated experiment was performed with a luminosity of  
$N_\mu \, m_\mathrm{kt} = 2\cdot 10^{22} ~\,\mathrm{kt}\,
\mathrm{year}$, which is by a factor ten higher than the standard 
value used in this work. It can be seen that despite the relatively 
high value of $N_\mu \, m_\mathrm{kt}$, the results are not good. 
Particularly, the error ellipses grow drastically with
decreasing $\theta_{12}$ and $\dm{21}$.
%
%%%%%%%%%%%%%%%%%%%%%%%%%%%%%%%%%%%%%%%%%%%%%%%%%%%%%%%%%%%%%%%%%%%%%
\subsection*{\boldmath Correlation of $\deltacp$ with $\theta_{13}$ 
and possible degeneracies}
\label{sec:th13cpcorr}
%%%%%%%%%%%%%%%%%%%%%%%%%%%%%%%%%%%%%%%%%%%%%%%%%%%%%%%%%%%%%%%%%%%%%

In section \ref{sec:SUBLEADING} it was demonstrated that the
parameters $\theta_{13}$ and $\deltacp$ can be strongly 
correlated. This effect was visualized in the right plot
of fig.~\ref{fig:errorbarsTh13}. It is
obviously important for studies of the CP-phase $\deltacp$
and it is discussed in detail in ref.~\cite{Burguet-Castell:2001ez}. 
Since our statistical method takes into account all possible two 
parameter correlations, also this particular correlation 
is automatically included in our results. We do not further
discuss this correlation but focus on another interesting
problem which is also studied in the above referred work. 
\begin{figure}[htb!]
\begin{center}
\includegraphics[width=0.4\textwidth,angle=-90]{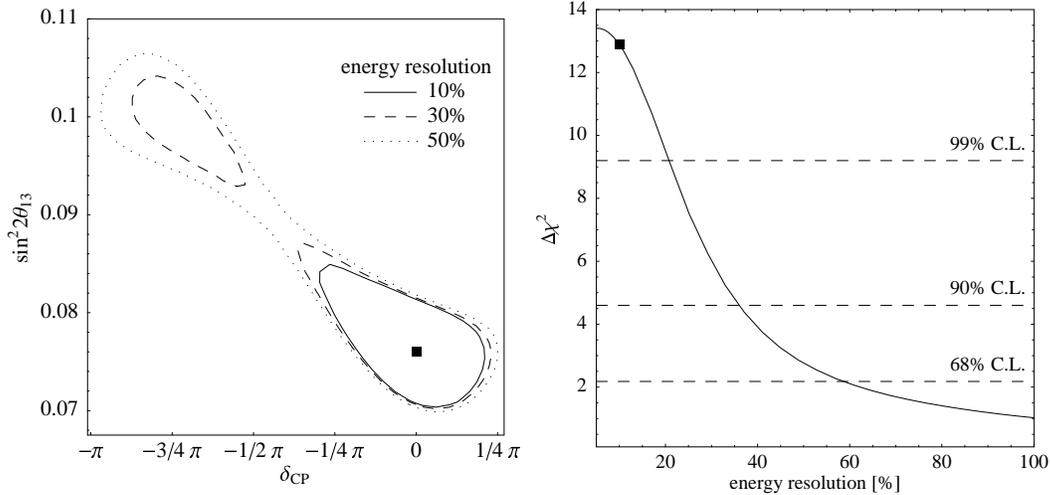}
\end{center}
\caption{Fits to $\sin^2 2\theta_{13}$ and
$\deltacp$ with different values of the energy resolution of the
detector (left plot) and $\chi^2$--difference between the best fit
points of the two degenerate solutions as function of
the energy resolution (right plot) . The black square in the right 
plot denotes the standard energy resolution used in this work. 
Parameters: $E_\mu=50\GeV$, $L=3000\,\mathrm{km}$, 
$\dm{21}=10^{-4}\thinspace\mathrm{eV}^2$.}
\label{fig:degen}
\end{figure}
There, it is demonstrated that considering the full parameter 
space of $\deltacp$, multiple degenerate solutions appear in
simultaneous fits of $\deltacp$ and $\theta_{13}$.
It is also stated that this degeneracy could possibly be 
lifted when different baselines or beam energies are studied 
simultaneously. We find that the appearance of degenerate solutions
crucially depends on the energy resolution of the detector. 
In fig.~\ref{fig:degen} the influence of the energy 
resolution on a fit of $\deltacp$ versus $\sin^2 2\theta_{13}$ 
is shown. In the left hand plot the fit itself is depicted for 
the three energy resolutions 10\%, 30\% and 50\%. The 
second degenerate solution in the upper part of the plot 
diminishes with increasing resolution and completely vanishes
for the value 10\%. To further illustrate this effect, 
the right hand plot shows the 
$\chi^2$-difference of the best fit points of the two degenerate 
solutions as function of the energy resolution of the detector. 
The dashed horizontal lines in the plot indicate which energy 
resolution is needed to refuse the second degenerate solution
at a certain confidence level. With an energy resolution of 
10\%, like we use it throughout this study\footnote{This 
particular value seems reasonable since it is very close to 
the value which is quoted by the MONOLITH 
collaboration~\cite{Agafonova:2000xm}. Details to the 
definition of energy resolution can be found in 
section~\ref{sec:numerics}.}, the second solution disappears 
with a confidence of more than 99\%. We checked that 
this holds for all values of $\deltacp$ and $\theta_{13}$. 
In ref.~\cite{Burguet-Castell:2001ez} the analysis was 
performed with only five energy bins. This corresponds to
an energy resolution, which is too bad to allow a lift
of the degeneracy.

%%%%%%%%%%%%%%%%%%%%%%%%%%%%%%%%%%%%%%%%%%%%%%%%%%%%%%%%%%%%%%%%%%%%%
\subsubsection*{\boldmath Optimization of baseline $L$ and muon energy $E_\mu$}
\label{sec:optimalLE}
%%%%%%%%%%%%%%%%%%%%%%%%%%%%%%%%%%%%%%%%%%%%%%%%%%%%%%%%%%%%%%%%%%%%%
%
\begin{figure}[htb!]
\begin{center}
\includegraphics[width=\textwidth,angle=0]{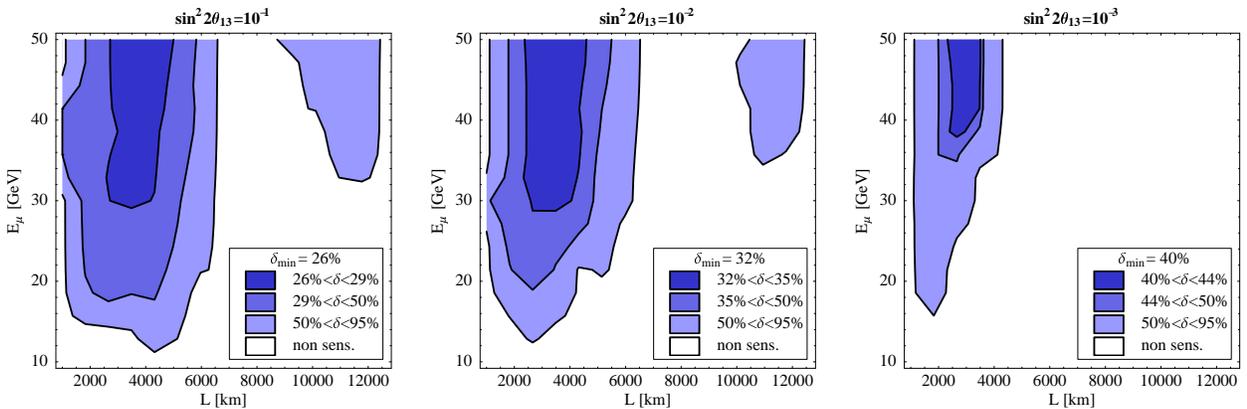}
\end{center}
\caption{Results of fits of the CP phase $\deltacp$
as function of the baseline $L$ and the muon energy $E_\mu$ 
for $\dm{31}=3.5\cdot 10^{-3}\eV^2$, $\dm{21} = 10^{-4} \eV^2$, 
$\theta_{23}=\pi/4$, $\theta_{12}=\pi/4$, $N_\mu \, m_\mathrm{kt} = 
2\cdot 10^{21} \,\mathrm{kt}\,\mathrm{year}$ and three values 
of $\sin^2 2\theta_{13}$ ($10^{-1}$, $10^{-2}$, $10^{-3}$). 
Dark shading indicates the preferred regions. The quantity 
$\delta$ plotted here is the percentage of the $\deltacp$ 
parameter space $[-\pi/2,\pi/2]$ which is compatible with 
the simulated experimental data at the $3\sigma$ confidence 
level. The contour lines correspond to $\delta=50\%$ and 
$\delta= 95\%$. In the white shaded region no information 
on the CP phase can be obtained.}
\label{fig:LEoptSubsubleading}
\end{figure}
In fig.~\ref{fig:LEoptSubsubleading} the regions in 
$L$-$E_\mu$-plane are shown where $\deltacp$ can be measured. 
We find that in general higher energies and baselines around 
$3000\,\mathrm{km}$ are preferred which is in good agreement 
with the results obtained in~\cite{CERVERA,Burguet-Castell:2001ez} 
but is in clear contradiction to the results obtained 
in~\cite{Koike:2000jf,Koike:2001kv}. The reason for this 
disagreement seems to lie in the different use of statistics. 
In the referred work no fits are performed but Gaussian error 
propagation is used (see sec.~\ref{sec:statistics}). 
In table~\ref{tab:dcp} the 
optimal regions in the $L$-$E_\mu$-plane are given for different 
values of $\sin^22\theta_{13}$.
\begin{table}[ht!]
\begin{center}
\begin{tabular}{|c|c||c|c|}
\hline
$\dm{31}$ & $\sin^2 2\theta_{13}$ & baseline & beam energy \\ \hline \hline
& $10^{-1}$ & $2400\,\mathrm{km} \lesssim L \lesssim 4200\,\mathrm{km}$ & 
$E_\mu \gtrsim 40\GeV$ \\ \cline{2-4} 
$6.0\cdot 10^{-3}\eV^2$ & $10^{-2}$ & 
$2400\,\mathrm{km} \lesssim L \lesssim 4000\,\mathrm{km}$ & 
$E_\mu \gtrsim 40\GeV$ \\ \cline{2-4} 
& $10^{-3}$ & $1800\,\mathrm{km} \lesssim L \lesssim 2700\,\mathrm{km}$ & 
$E_\mu \gtrsim 45\GeV$ \\ 
\hline\hline
& $10^{-1}$ & $2800\,\mathrm{km} \lesssim L \lesssim 4500\,\mathrm{km}$ & 
$E_\mu \gtrsim 30\GeV$ \\ \cline{2-4} 
$3.5\cdot 10^{-3}\eV^2$ & $10^{-2}$ & 
$2500\,\mathrm{km} \lesssim L \lesssim 4500\,\mathrm{km}$ & 
$E_\mu \gtrsim 30\GeV$ \\ \cline{2-4} 
& $10^{-3}$ & $2500\,\mathrm{km} \lesssim L \lesssim 3500\,\mathrm{km}$ & 
$E_\mu \gtrsim 40\GeV$ \\ 
\hline\hline
& $10^{-1}$ & no sens. &  no sens. \\ \cline{2-4} 
$1.0\cdot 10^{-3}\eV^2$ & $10^{-2}$ & no sens. & no sens. \\ \cline{2-4} 
& $10^{-3}$ &  no sens. &  no sens. \\ \hline
\end{tabular}
\caption{\label{tab:dcp} Optimal choice of the baseline $L$ 
and the beam energy $E_\mu$ for measurements of the CP phase 
$\deltacp$. The data of the first three lines are taken from 
figs.~\ref{fig:LEoptSubsubleading}. The given ranges identify 
the regions where the statistical error $\delta(\deltacp)$ is 
not more than by a factor 1.1 larger than the optimal value 
$\delta(\deltacp)_\mathrm{min}$. In figs.~\ref{fig:LEoptSubsubleading} 
these regions are indicated by the darkest shading.
}
\end{center}
\end{table}
We also checked the influence of the value of $\dm{31}$ on the 
sensitivity and the optimal regions in the  $L$-$E_\mu$-plane
(see table~\ref{tab:dcpdm31} ). The sensitivity gets better (worse) 
as $\dm{31}$ increases (decreases). The optimal baseline seems to be quite 
stable against changes of $\dm{31}$. The optimal energy however gets 
smaller as $\dm{31}$ gets smaller. 
\begin{table}[ht!]
\begin{center}
\begin{tabular}{|c|c||c|}
\hline
$\dm{31}$ & $\sin^2 2\theta_{13}$ & fraction \\ \hline \hline
& $10^{-1}$ & 22\% \\ \cline{2-3} 
$6.0\cdot 10^{-3}\eV^2$ & $10^{-2}$ & 24\% \\ \cline{2-3} 
& $10^{-3}$ & 28\% \\ 
\hline\hline
& $10^{-1}$ &  26\% \\ \cline{2-3} 
$3.5\cdot 10^{-3}\eV^2$ & $10^{-2}$ & 32\% \\ \cline{2-3} 
& $10^{-3}$ & 40\% \\ 
\hline\hline
& $10^{-1}$ & 100\% \\ \cline{2-3} 
$1.0\cdot 10^{-3}\eV^2$ & $10^{-2}$ &100\% \\ \cline{2-3} 
& $10^{-3}$ & 100\% \\ \hline
\end{tabular}
\caption{\label{tab:dcpdm31} Fraction of the parameter space of 
$\deltacp$ which is covered by the $3\sigma$ acceptance region. 
100\% means that no information on the CP phase can be extracted 
from the experimental data. The values are based on $\dm{21} = 
10^{-4} \eV^2$, $\theta_{23}=\pi/4$, $\theta_{12}=\pi/4$ and
$N_\mu \, m_\mathrm{kt} = 2\cdot 10^{21} \,\mathrm{kt}\,\mathrm{year}$.}
\end{center}
\end{table}
%

%%%%%%%%%%%%%%%%%%%%%%%%%%%%%%%%%%%%%%%%%%%%%%%%%%%%%%%%%%%%%%%%%%%%%
\subsection*{\boldmath Sensitivity reach for $\deltacp$}
\label{sec:cpreach}
%%%%%%%%%%%%%%%%%%%%%%%%%%%%%%%%%%%%%%%%%%%%%%%%%%%%%%%%%%%%%%%%%%%%%

The two parameters $\dm{21}$ and
$\theta_{13}$ control the size of all CP-effects. A very important 
question thus is, down to which values of these two parameters 
the CP phase can be determined.
\begin{figure}[htb!]
\begin{center}
\includegraphics[width=0.5\textwidth,angle=0]{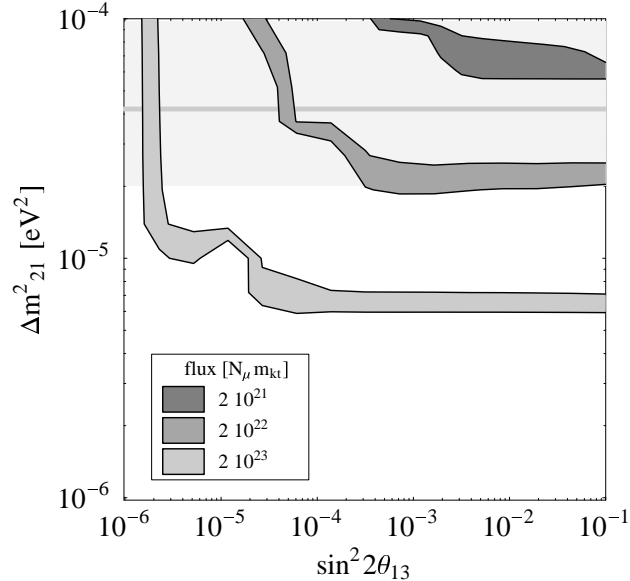}
\end{center}
\caption{Sensitivity limit of measurements of the CP-phase $\deltacp$
in the $\theta_{13}$-$\dm{21}$ parameter plane for three different values 
of $N_\mu \, m_\mathrm{kt}$.
The lower edge of each band represents the sensitivity limit
under which no information on the CP-phase can be extracted from the
experimental data. The upper edges of each band indicates  
50\% statistical error (at 99\%~C.L.) for measurements of $\deltacp$.
This result was obtained with a baseline of $3000\,\mathrm{km}$. 
The light grey shaded region in the background indicates the values
of $\dm{21}$ compatible with the LMA-MSW solution and the grey horizontal 
line indicates its best fit value.}
\label{fig:CPsens}
\end{figure}
This sensitivity reach is shown in fig.~\ref{fig:CPsens}. The three
grey bands were obtained with luminosities  $N_\mu \, m_\mathrm{kt}$ of  
$2\cdot 10^{21} \,\mathrm{kt}\,\mathrm{year}$ (dark grey), 
$2\cdot 10^{22} \,\mathrm{kt}\,\mathrm{year}$ (grey) and 
$2\cdot 10^{23} \,\mathrm{kt}\,\mathrm{year}$ (light grey).
The upper edge of each band represents 50\% error, which is 
the contour line where half of the parameter space of $\deltacp$ can be
ruled out by the experiment.
The lower edge represents the sensitivity limit under which no information
on $\deltacp$ can be extracted from the experimental data. 
The light grey shaded area in the background indicates the presently allowed 
LMA-MSW solution and the grey horizontal line 
its best fit value\cite{Bahcall:2001hv}. 
Note that the so called ``initial stage'' option $2\cdot 10^{20} \,\mathrm{kt}
\,\mathrm{year}$ is not included in the plot since it does not provide 
sufficient event rates to make the CP-effects accessible. 
With the standard luminosity assumed in this work ($2\cdot 10^{21} \,\mathrm{kt}
\,\mathrm{year}$), CP-effects are only accessible for values of $\dm{21}$ at 
the upper edge of the LMA-solution. If $\dm{21}$ is at the lower edge, 
the luminosity must be increased by a factor of more than 10. In this case, 
however, systematic errors could become the dominant source of uncertainties.

%%%%%%%%%%%%%%%%%%%%%%%%%%%%%%%%%%%%%%%%%%%%%%%%%%%%%%%%%%%%%%%%%%%%%%
\section{Conclusion}
\label{sec:conclusion}
%%%%%%%%%%%%%%%%%%%%%%%%%%%%%%%%%%%%%%%%%%%%%%%%%%%%%%%%%%%%%%%%%%%%%

Previous neutrino factory studies were based on simplified 
treatments, which do not include all parameter correlations. 
We show that these correlations have considerable influence 
on the physics potential of a neutrino factory, especially 
for measurements close to the sensitivity limit of 
$\theta_{13}$ and for measurements of CP violation. 
We developed for our analysis improved statistical methods
which are based on the computation of all two parameter 
correlations and which automatically include all parameter 
uncertainties. We used this method to look for all possible
correlations and to refine in this way the understanding of the 
potential of neutrino factories. Using this method we found 
indeed correlations which were previously ignored or
overlooked, leading to corrections for earlier results.
The results which are summarized below were obtained with a 
default neutrino flux of $2\cdot 10^{20}$ decaying
muons per year directed to a $10\,$kt magnetized iron detector.
Systematical, experimental and background limitations 
are not included in our study and the given errors and 
limits are thus of pure statistical nature. Unless mentioned 
differently the central value of the parameter $\dm{31}$ was 
chosen to be $3.5\cdot 10^{-3}\eV^2$.

We found that the minimal errors which can be achieved for the
leading parameters $\theta_{23}$ and $\dm{31}$ are roughly 8\% 
and 6\%, respectively. The precise length of the baseline is 
not crucial for these measurements, as long as it is larger 
than a certain minimal value. This lower limit depends strongly on 
$\dm{31}$. For $\dm{31} = 3.5\cdot 10^{-3}\eV^2$ a baseline
of $3000\,$km is enough to obtain the precision mentioned above.
Lower values of $\dm{31}$ required larger baselines. For  
$\dm{31} = 1.0\cdot 10^{-3}\eV^2$ the baseline should, for 
example, exceed $5000\,$km to obtain optimal results and the 
precision which can be achieved is only 20\%. 
$\dm{31}$ values above $ 3.5\cdot 10^{-3}\eV^2$ allow on the
other hand correspondingly shorter baselines. High beam 
energies are in general better for measurements of $\theta_{23}$ 
and $\dm{31}$, but the energy dependence becomes rather weak
between $30\GeV$ and $50\GeV$. Correlations with the \Psl~parameters 
$\dm{21}$ and $\theta_{12}$ become for large $\dm{21}$ important
and contribute significantly to the total errors of $\theta_{23}$ 
and $\dm{31}$. A potential measurement of $\dm{21}$ in the LMA-MSW
regime by KamLAND will therefore improve the errors on $\dm{31}$ 
by up to a factor of three.

For measurements of the mixing angle $\theta_{13}$, we also 
found that beam energies between $30\GeV$ and $50\GeV$ perform
approximately equally well, but two cases should be 
distinguished for the baseline. For $\theta_{13}$ somewhat 
below the current upper bound baselines between $3000\,$km 
and $8000\,$km are recommended. If, however, $\theta_{13}$ 
is very small and therefore close to the sensitivity limit, 
we find a preferred baselines between $7000\,$km and $8000\,$km. 
The reason behind this is that there exists a quiet strong 
correlation between $\theta_{13}$ and the \Pssl~parameters 
$\dm{21}$, $\theta_{12}$ and $\deltacp$. This correlation is 
important for large values of $\dm{21}$ but it looses in 
strength for larger baselines. Our improved statistical 
analysis gives sensitivity limits for measurements of $\theta_{13}$
which are up to one order of magnitude worse than previous
results. If the LMA-MSW region is confirmed, then KamLAND's
measurement of $\dm{21}$ will improve the determination of 
$\theta_{13}$ at a neutrino factory for $\cos\deltacp > 0$.
We found that the sensitivity limit for $\theta_{13}$ 
lies between $\sin^2 2\theta_{13} = 10^{-4}$ 
and $\sin^2 2\theta_{13} = 10^{-3}$, depending on the value of 
$\dm{21}$ and on the baseline. A 10\% statistical error on 
$\theta_{13}$ is expected if $\theta_{13}$ is close to the 
CHOOZ bound. A 50\% error is still achievable for 
$\sin^2 2\theta_{13} = 10^{-2}$. For a discussion of the
flux dependence we refer to the corresponding sections
of this work. We did not discuss in detail measurements of 
the sign of $\dm{31}$ and the prove of the MSW-effect 
since these points are closely related to measurements 
of $\theta_{13}$. One can thus roughly identify the 
sensitivity limits for measurements of $\theta_{13}$
with the limits down to which MSW-effects and the sign of 
$\dm{31}$ are measurable.

Finally, we focused on the parameters $\dm{21}$, $\theta_{12}$ 
and $\deltacp$. In the LMA-MSW case a neutrino factory can 
also measure $\dm{21}$ and $\theta_{12}$, but we have shown 
that a neutrino factory can not compete with KamLAND.
To improve this situation, fluxes at least a factor 100 
larger than usually discussed in the context of neutrino
factories are necessary. The reason 
for this is that the appearance probabilities only depend on 
the product $\dm{21}\sin2\theta_{12}$, which makes it difficult 
to obtain separately information on $\dm{21}$ and $\theta_{12}$.
Concerning measurements of the CP-phase $\deltacp$, we obtained 
the following results: Baselines between $2800\,$km and $3500\,$km 
are good choices. Lower baselines are not recommended, not only 
because of backgrounds which would spoil the signal, but also 
from a statistical point of view. Muon energies between
$30\GeV$ and $50\GeV$ are again preferred for such measurements. 
We find, however, that $N_\mu \, m_\mathrm{kt} = 2\cdot 10^{21} \,
\mathrm{kt}\,\mathrm{year}$ is lower bound for a CP violation 
measurement, which is at this limit only possible for very 
large values of $\dm{21}$. To cover the full LMA-MSW region, a 
luminosity of at least $N_\mu\,m_\mathrm{kt} = 
2\cdot 10^{22}\,\mathrm{kt}\,\mathrm{year}$ is needed. 
Moreover, we found that the degeneracy in the 
$\theta_{13}$-$\deltacp$ parameter plane, which was recently 
pointed out in the literature is not present in our analysis. 
Such a degeneracy shows only up when the energy resolution of 
the detector is reduced. 
 
We would like to stress again, that we perform a statistical analysis
and that systematic errors and backgrounds are not included.
The adopted statistical method resembles to a large extent the 
results which would be obtained by full six parameter fits. 
Nevertheless, the errors are in some cases still somewhat 
underestimated and we found that full six parameter fits can 
have up to a factor two larger errors. 

%%%%%%%%%%%%%%%%%%%%%%%%%%%%%%%%%%%%%%%%%%%%%%%%%%%%%%%%%%%%%%%%%%%%
%%%%                      Acknowledgments                      %%%%
%%%%%%%%%%%%%%%%%%%%%%%%%%%%%%%%%%%%%%%%%%%%%%%%%%%%%%%%%%%%%%%%%%%%

\vspace*{7mm}
Acknowledgments: We wish to thank E.~Akhmedov and T.~Ohlsson for 
discussions and useful comments.

%%%%%%%%%%%%%%%%%%%%%%%%%%%%%%%%%%%%%%%%%%%%%%%%%%%%%%%%%%%%%%%%%%%%%
%%%%                       References                            %%%%
%%%%%%%%%%%%%%%%%%%%%%%%%%%%%%%%%%%%%%%%%%%%%%%%%%%%%%%%%%%%%%%%%%%%%

\newpage
\bibliographystyle{./apsrev}

\bibliography{./paper}

%%%%%%%%%%%%%%%%%%%%%%%%%%%%%%%%%%%%%%%%%%%%%%%%%%%%%%%%%%%%%%%%%%%%%
\end{document}